\documentclass[letterpaper,journal]{IEEEtran}
\usepackage{amsmath,amsfonts}
\usepackage{array}
\usepackage{cite}
\usepackage{textcomp}
\usepackage{stfloats}
\usepackage{url}
\usepackage{verbatim}
\usepackage{orcidlink}
\usepackage{graphicx}
\usepackage[utf8]{inputenc} 
\DeclareUnicodeCharacter{FF0C}{,} 
\usepackage{graphicx}
\usepackage{textcomp}
\usepackage{xcolor}
\usepackage{multirow} 
\usepackage{subfigure}
\usepackage{bm}
\usepackage{verbatim}
\usepackage{booktabs}
\usepackage{balance}
\usepackage{endnotes}
\usepackage{bbding}
\usepackage{makecell}

\usepackage{url}
\usepackage{booktabs}
\usepackage{enumitem}
\usepackage{makecell}
\usepackage{xcolor}
\usepackage{graphicx}
\usepackage[export]{adjustbox}
\usepackage{xspace}
\usepackage{multirow}
\usepackage{balance}
\usepackage{amssymb}

\usepackage{bbm}
\usepackage{amsfonts}

\usepackage[linesnumbered,ruled,vlined]{algorithm2e} 
\usepackage{algpseudocode}
\usepackage{amsmath}

\usepackage{comment}
\usepackage{amsmath,amsfonts}
\usepackage{graphicx}
\usepackage{textcomp}
\usepackage{xcolor}
\usepackage{url}
\usepackage{amsmath}
\usepackage{makecell} 
\usepackage{subcaption}
\usepackage{hyperref}
\hypersetup{
    colorlinks=true,
    linkcolor=blue,
    citecolor=blue,
    urlcolor=blue,
    pdfborder={0 0 0}
}
\usepackage{times}
\usepackage{soul}
\usepackage{enumitem}
\usepackage{multirow}
\usepackage{bbm}
\usepackage{xspace}
\usepackage{url}
\usepackage{xcolor}
\usepackage[linesnumbered,ruled,vlined]{algorithm2e}

\usepackage{amsthm}
\SetKwInput{KwInput}{Input}
\SetKwInput{KwOutput}{Output}

\newtheoremstyle{examplestyle} 
    {1pt} 
    {1pt} 
    {\itshape} 
    {} 
    {\bfseries} 
    {.} 
    {.1em} 
    {} 
\theoremstyle{examplestyle}
\newtheorem{example}{Example}[section]

\newtheoremstyle{definitionstyle} 
    {1pt} 
    {1pt} 
    {\itshape} 
    {} 
    {\bfseries} 
    {.} 
    {.1em} 
    {} 
\theoremstyle{definitionstyle}
\newtheorem{definition}{Definition}[section]

\newcommand{\bluecomment}[1]{\textcolor{blue}{\texttt{/* #1 */}}}

\makeatletter
\newcommand{\unbalance}{\let\@dblfloatplacement\@dbldonotplace}
\makeatother
\def\BibTeX{{\rm B\kern-.05em{\sc i\kern-.025em b}\kern-.08em
    T\kern-.1667em\lower.7ex\hbox{E}\kern-.125emX}}
\usepackage{balance}
\begin{document}
\title{Advancing Knowledge Tracing by Exploring Follow-up Performance Trends}

\author{Hengyu Liu\orcidlink{0000-0001-6545-7181}, Yushuai Li\orcidlink{0000-0002-3043-3777}, Minghe Yu\orcidlink{0000-0002-0287-8867}, Tiancheng Zhang\orcidlink{0000-0001-6902-9299}, Ge Yu\orcidlink{0000-0002-3171-8889}, Torben Bach Pedersen\orcidlink{0000-0002-1615-777X}, Kristian Torp\orcidlink{0000-0002-8239-0262}, Christian S. Jensen\orcidlink{0000-0002-9697-7670}, Tianyi Li\orcidlink{0000-0001-5424-6442}
\thanks{Hengyu Liu, Yushuai Li, Torben Bach Pedersen, Kristian Torp, Christian S. Jensen, and Tianyi Li are with Aalborg University, Aalborg 9220, Denmark. (email: \{heli, yusli,  tbp, torp, csj, tianyi\}@cs.aau.dk).
Minghe Yu, Tiancheng Zhang, and Ge Yu are with Northeastern University, Shenyang, China. (email: \{yuminghe,tczhang,yuge\}@neu.edu.cn).
}
}

\markboth{IEEE TRANSACTIONS ON KNOWLEDGE AND DATA ENGINEERING， VOL.XX, NO.X, XXXX}%
{How to Use the IEEEtran \LaTeX \ Templates}

\maketitle

\begin{abstract}
Intelligent Tutoring Systems (ITS), such as Massive Open Online Courses, offer new opportunities for human learning. At the core of such systems, knowledge tracing (KT) predicts students' future performance by analyzing their historical learning activities, enabling an accurate evaluation of students' knowledge states over time. 
We show that existing KT methods often encounter \textbf{correlation conflicts} when analyzing the relationships between historical learning sequences and future performance.
To address such conflicts, we propose to extract so-called Follow-up Performance Trends (FPTs) from historical ITS data and to incorporate them into KT.
We propose a method called \textbf{\underline{F}}orward-Look\textbf{\underline{i}}ng K\textbf{\underline{n}}owledg\textbf{\underline{e}} T\textbf{\underline{r}}acing (\textsf{FINER}) that combines historical learning sequences with FPTs to enhance student performance prediction accuracy. 
\textsf{FINER} constructs learning patterns that facilitate the retrieval of FPTs from historical ITS data in linear time; 
\textsf{FINER} includes a novel similarity-aware attention mechanism that aggregates FPTs based on both frequency and contextual similarity; 
and \textsf{FINER} offers means of combining FPTs and historical learning sequences to enable more accurate prediction of student future performance. 
Experiments on six real-world datasets show that \textsf{FINER} can outperform ten state-of-the-art KT methods, increasing accuracy by 8.74\% to 84.85\%.
The source code and implementation details of \textsf{FINER} are publicly available\footnote{\href{https://github.com/hyLiu1994/FINER}{\textcolor{blue}{https://github.com/hyLiu1994/FINER}}}.
\end{abstract}

\section{Introduction}\label{sec:intro}
\IEEEPARstart{I}{ntelligent}
tutoring systems (ITS), which encompass Massive Open Online Courses \cite{DBLP:conf/sigir/Gong0WFP0Y20} and Online Judging systems \cite{DBLP:conf/dasfaa/DongHL20}, offer novel opportunities for independent and effective student learning. In ITS, knowledge tracing (KT) is employed to model the knowledge states of students over time. Specifically, KT predicts the future performance of students based on their engagement with exercises \cite{DBLP:conf/nips/PiechBHGSGS15,DBLP:conf/aaai/VieK19}, which is important for enabling effective learning. 

Existing KT methods~\cite{Zhang2018,Liu2022,long_improving_2022,DBLP:conf/aaai/00060HL023,DBLP:conf/cikm/PandeyS20,huang2023towards,wang_perm_2022, kim_leveraging_2023,zhang2024coskt} primarily analyze historical learning sequences by focusing on performance patterns in identical or similar questions \cite{DBLP:conf/nips/PiechBHGSGS15,ghosh2020context,DBLP:conf/edm/PandeyK19,Gorgun2022}, while giving more weight to recent learning behaviors \cite{shen2021learning,DBLP:conf/edm/ChoffinPBV19, xu2023learning}. 
Although this targeted approach helps maintain computational efficiency and prevents overfitting, it overlooks broader learning patterns that may provide valuable context, making these methods susceptible to \textbf{correlation conflicts}.

\begin{figure}
\centering
\includegraphics[width=0.5\textwidth]{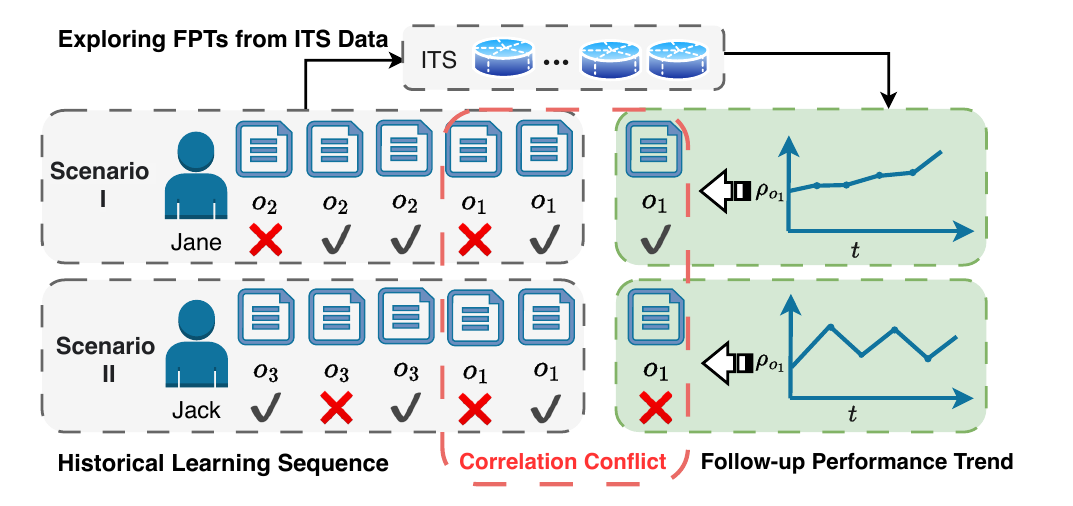}
\caption{Motivating example.} 
\label{fig:sec1-toyexample}
\end{figure}

\begin{example}~\label{ex:motivation1}
Consider three questions $o_1$, $o_2$, and $o_3$, each of which has multiple problem-solving strategies (e.g., linear programming problem \cite{sole2016multiple}). 
We examine two learning sequences from students Jane and Jack under different scenarios (illustrated in Fig.~\ref{fig:sec1-toyexample}).
\textbf{In Scenario I, students try to practice on more problems and only try one strategy for each question.} Thus, when Jane encounters $o_1$ again after answering it correctly, she answers it correctly since she has mastered one strategy of $o_1$. 
\textbf{In Scenario II, students may practice different strategies for the same question}. Thus, Jack answers $o_1$ wrongly after having answered it correctly because he is attempting a different strategy of $o_1$.
\end{example}

As shown by the red dashed box in Fig.~\ref{fig:sec1-toyexample}, by focusing primarily on identical or similar questions and recent behavior (e.g., $(o_1,  \times),  (o_1,  \checkmark)\rangle$), existing KT models struggle to differentiate scenarios where \textbf{identical learning patterns lead to divergent performance outcomes on the same question}. We refer to this phenomenon as \textbf{correlation conflict}, which notably limits model effectiveness. Our analysis (see Section~\ref{sec:analysis-correlation-conflict} and Table~\ref{tab:correlation-conflict-new}) reveals that such correlation conflicts are prevalent, occurring in 5.20\%--10.15\% or more cases across six widely used datasets. 

To address \textbf{correlation conflicts} while preserving the computational efficiency and generalization capabilities of existing KT methods, we propose to integrate Follow-up Performance Trends (FPTs) into KT. FPTs represent performance over time following the current learning pattern, as shown on the right side of Fig.~\ref{fig:sec1-toyexample}, which can be fetched from students' historical learning sequences in the ITS according to their learning pattern. 

\begin{example}~\label{ex:motivation1-2} 
Continuing Example~\ref{ex:motivation1}, extract Jane and Jack’s learning patterns $v_{\mathit{Jane}} = \langle (o_2, \times), (o_2,  \checkmark), (o_2, \checkmark) , (o_1, \times), \\ (o_1, \checkmark)\rangle$, $v_{\mathit{Jack}} = \langle (o_3, \checkmark),  (o_3,  \times), (o_3,  \checkmark), (o_1,  \times), ( o_1,  \checkmark)\rangle$. These learning patterns are suffixes of Jane and Jack's historical learning sequence. Next, we retrieve $\textit{FPT}_{\mathit{Jane}}$, $\textit{FPT}_{\mathit{Jack}}$ from historical ITS data, where each $\textit{FPT}$ captures the change in the probability of correctly answering $o_1$ over time. 
$\textit{FPT}_{\mathit{Jane}}$ shows a gradual improvement in answering $o_1$ after $v_{\mathit{Jane}}$. Correspondingly, $\textit{FPT}_{\mathit{Jack}}$ shows fluctuations. These distinct FPTs effectively characterize the different scenarios, enabling KT models to differentiate between them and resolve correlation conflicts.
\end{example}

Example~\ref{ex:motivation1-2} illustrates the role of FPTs in resolving correlation conflicts and enhancing student performance prediction accuracy. However, integrating FPTs into KT is non-trivial.

\noindent{\textit{\textbf{Challenge I: Which historical data should be used to formulate FPTs, and how to support efficient retrieval of FPTs?}}} 
As this is the first study on using FPTs for enhancing performance prediction, a key step is to identify the ITS data that is most relevant when forming FPTs. Moreover, FPTs vary over time as they mirror learning behaviors that evolve. How to extract FPTs from extensive historical ITS data to enable real-time prediction of student performance is challenging. 
Existing pattern retrieval proposals \cite{knuth1977fast,yao1979complexity,choi2007pattern,kociumaka2019pattern,anggreani2020knuth,sha2022chaotic} do not offer real-time performance. 

\noindent{\textit{\textbf{Challenge II: How to determine confidence scores of FPTs?}}} 
Learning patterns can be any suffixes of the student's recent learning sequence, and different learning patterns may correspond to different FPTs. Thus, to incorporate a broad range of information, we extract FPTs w.r.t. student learning patterns of varying lengths, thereby enhancing prediction accuracy. However, this hinges on our ability to assign confidence scores to FPTs. Simply computing scores based on frequencies in historical learning sequences may disregard less frequent but equally important information \cite{calandriello2022scaling,host2019data}. Thus, determining the confidence of an FPT is an open challenge.

\noindent{\textit{\textbf{Challenge III: How to effectively integrate FPTs with historical learning sequences?}}} 
Existing methods~\cite{piccialli2021artificial,wilson2023calda,li2022metro} typically merge multi-source time-series data sequentially due to their temporal overlap. This approach becomes problematic when combining FPTs with historical sequences. 
Continuing Example~\ref{ex:motivation1-2}, we have probabilities $\langle \rho_{o_1}[1], \rho_{o_1}[2] \rangle$ from the FPT, indicating the likelihood of correctly answering $o_1$ in the next two attempts, and Jane's most recent learning pattern $\langle (o_3, \checkmark), (o_1, \times) \rangle$. 
It is difficult to establish the temporal alignment of probabilities and actual learning patterns. Temporal misalignment makes it challenging to effectively fuse FPTs with historical learning sequences. 

We propose a method called \textbf{\underline{F}}orward-Look\textbf{\underline{i}}ng K\textbf{\underline{n}}owledg\textbf{\underline{e}} T\textbf{\underline{r}}acing (\textsf{FINER}) that effectively integrates student FPTs with corresponding historical learning sequences with the aim of enabling more accurate prediction of student learning performance. \textsf{FINER} comprises three modules. 
Addressing Challenges I, an FPT Search Module extracts learning patterns to build a learning pattern trie from the ITS data. The trie, coupled with novel algorithms, enables pattern location and FPT retrieval in linear time. 
To address Challenge II, we incorporate a similarity-aware attention mechanism into an FPT Aggregation Module. The mechanism assigns similar FPTs to similar confidence scores, even if their frequencies vary. The idea is to value the quality of FPTs over sheer quantity, which recognizes that infrequent patterns can be as revealing as frequent ones. The module then aggregates FPTs based on their confidence scores. 
In response to Challenge III, we propose a Recent History Fusion Module that independently encodes historical sequences and FPTs to avoid direct temporal alignment, fuses their representations through tensor outer products, and models temporal dependencies at the feature level using LSTM networks.
The contributions are summarized as follows.

\setlength{\parsep}{0pt}
\begin{itemize}[itemsep=1pt, leftmargin=10pt]
\item To the best of our knowledge,  \textsf{FINER} is the first method to integrate FPTs with student historical learning sequences, which allows us to solve correlation conflicts.

\item We propose a learning pattern trie that stores historical data relevant to FPTs. Along with novel algorithms, this trie enables retrieval of FPTs in linear time. 

\item We propose a novel similarity-aware attention mechanism to aggregate FPTs corresponding to learning patterns of varying lengths. We enable fusion of FPTs with historical learning sequences by proposing a Historical-FPT Fusion Network, making it possible to determine future performance of students accurately and efficiently.

\item Experiments on six real-world datasets offer evidence that \textsf{FINER} outperforms state-of-the-art KT models, increasing accuracy by {8.74}\% to {84.85}\% and improving efficiency by {1.39}\% to {4.11}\%. 
\end{itemize}

The paper is organized as follows. Section~\ref{sec: related work} reviews related work. Section~\ref{sec:PRELIMINARIES} presents the preliminaries, 
and Section~\ref{sec:FINER} provides the detailed design of \textsf{FINER}. Experimental results are reported in Section~\ref{sec: Experiments}, and Section~\ref{sec: Conclusion} concludes and offers research directions.

\section{Related Work} \label{sec: related work}
Knowledge Tracing is a foundational task in computational education that aims to model a student's evolving knowledge state over time by analyzing their historical interactions. An accurate KT model is the cornerstone of adaptive learning systems, enabling personalized feedback, optimal curriculum sequencing, and targeted interventions. The field has progressed from foundational statistical models to a diverse landscape of sophisticated deep learning architectures that address increasingly nuanced aspects of student learning.

\subsection{Foundational Approaches to Knowledge Tracing} 
The first generation of KT models was primarily statistical. The seminal work, Bayesian Knowledge Tracing (BKT) \cite{corbett1994knowledge}, employs a Hidden Markov Model where each skill is a latent binary variable (mastered or not mastered). While foundational, BKT's simplifying assumptions (e.g., one skill per question, knowledge being non-forgettable in its basic form) limit its applicability to complex, real-world learning scenarios. Zhang et al. \cite{Zhang2018} propose BKT with three learning states (true/unsure/false), extending the two-state (true/false) approach. Xu et al. \cite{xu2011using} develop a Logistic Regression-Dynamic Bayesian Network to determine transitional probabilities of knowledge in a dynamic Bayesian network. Recent Markov chain-based KT proposals~\cite{Liu2022,Gorgun2022} introduce fuzzy Bayesian methods to more effectively evaluate student cognitive performance in continuous scoring scenarios. Moreover, they emphasize the need for data pre-processing that enhances predictions by excluding disengaged responses. 

Another significant line of early research involves Factor Analysis methods~\cite{van2013handbook,cen2006learning, pavlik2009performance,DBLP:conf/aaai/VieK19}, which model performance by identifying latent factors like student ability and question difficulty.
Cen et al.~\cite{cen2006learning} propose a semi-automated method for improving a cognitive model that combines a statistical model, human expertise, and combinatorial search. 
Pavlik et al.~\cite{pavlik2009performance} improve the understanding of student performance by examining specific interactions with learning materials and their impact on knowledge acquisition. 
Vie et al.~\cite{DBLP:conf/aaai/VieK19} incorporate Factorization Machines into KT to model student learning by identifying complex patterns in student-tutor interactions.
Choffin et al. ~\cite{DBLP:conf/edm/ChoffinPBV19} integrate the temporal distribution of skill practicing and learning and forgetting curves for different skills to achieve enhanced performance in spaced repetition contexts. 
These models provided robust statistical frameworks but often require extensive feature engineering and struggle to capture the granular, temporal dynamics of learning as effectively as modern neural approaches.

\subsection{The Rise of Deep Learning in Knowledge Tracing}
The availability of large-scale educational datasets catalyzed the adoption of deep learning for KT, enabling models to automatically learn rich, hierarchical feature representations from raw interaction data.

\noindent \textbf{Sequential Models: From RNNs to Transformers.} The paradigm shift began with Deep Knowledge Tracing~\cite{DBLP:conf/nips/PiechBHGSGS15}, which first applied Recurrent Neural Networks~\cite{DBLP:conf/aaai/SuLLHYCDWH18,DBLP:conf/aaai/00060HL023} to model the student's interaction history as a sequence, treating the knowledge state as an evolving high-dimensional vector. 
This was followed by memory-augmented networks like the Dynamic Key-Value Memory Network~\cite{DBLP:conf/www/ZhangSKY17}, which uses an explicit memory component to better store and retrieve skill representations. The subsequent introduction of attention mechanisms~\cite{DBLP:conf/edm/PandeyK19,DBLP:conf/cikm/PandeyS20,ghosh2020context,huang2023towards,DBLP:conf/iclr/0001L0H023}, inspired by their success in natural language processing, allowed models to dynamically weigh the importance of past interactions. The Self-Attentive Knowledge Tracing~\cite{DBLP:conf/edm/PandeyK19} model was a pioneer in this space. This line of work culminated in highly effective Transformer-based architectures like Context-Aware Attentive Knowledge Tracing~\cite{ghosh2020context}, which introduced a more sophisticated monotonic attention mechanism that incorporates a learnable exponential decay term to model forgetting.  

\noindent \textbf{Graph-based Models for Relational Structures.} Recognizing that learning is not merely sequential, a significant recent trend~\cite{nakagawa2019graph,tong2020structure,tong2022introducing,wang2025knowledge,wang2024psychological} is the use of Graph Neural Networks (GNNs) to explicitly model the rich relational structures between students, questions, and skills. For instance, some models~\cite{wang2025knowledge} construct dual-graph convolutional networks to simultaneously capture student-student and skill-skill relationships, thereby alleviating data sparsity. Other innovative approaches like Psy-KT~\cite{wang2024psychological} build a heterogeneous graph of students, exercises, and skills, and uniquely incorporate psychological factors (e.g., frustration, concentration) to create a more holistic representation of the learning process.

\subsection{Modeling the Nuances of Human Learning}
Beyond predictive accuracy, modern KT research is increasingly focused on tackling more subtle challenges that are crucial for building robust and effective educational systems. 

\noindent \textbf{Modeling Forgetting and Context.} Models like KVFKT~\cite{guan2025kvfkt} explicitly integrate the Ebbinghaus forgetting curve, while AKT~\cite{ghosh2020context} learns a data-driven decay rate. The Psy-KT model also considers the Ebbinghaus curve in its framework~\cite{wang2024psychological}, and DKVMN\&MRI~\cite{xu2024dkvmn} introduces the Ebbinghaus function.

\noindent \textbf{Modeling Mistakes and Aberrant Behavior.} The concepts of ``slips'' and ``guesses'' are fundamental to KT~\cite{van2013handbook,cen2006learning}. Modern approaches tackle this with greater sophistication. Uncertainty-aware KT~\cite{cheng2025uncertainty} explicitly distinguishes between aleatory uncertainty (data noise like slips/guesses) and epistemic uncertainty (model uncertainty), using contrastive learning to become more robust. Another approach, Option Tracing (OT)~\cite{li2025integrating}, models the specific multiple-choice option a student selects, providing a richer signal about specific misconceptions.

\noindent \textbf{Improving Model Interpretability.} As models grow in complexity, their ``black-box'' nature becomes a barrier. A growing body of work~\cite{liu2020tracking,minn2022interpretable,DBLP:conf/aaai/00060HL023,sun2024interpretable,cui2024interpretable} focuses on creating interpretable models. For example, PSI-KT~\cite{zhou2024predictive} is a hierarchical generative model that achieves interpretability by design, explicitly modeling individual cognitive traits and the prerequisite structure of knowledge.


Despite these advancements, existing methods remain fundamentally \textbf{backward-looking}: they model a student's current state based on their past interactions. They may account for forgetting by decaying the influence of the past or for context by enriching the representation of a past event. However, they do not directly leverage information about what happens \textbf{after} a learning pattern occurs.

This is the critical gap addressed by \textsf{FINER}. To our knowledge, FINER is the first model to resolve the identified \textbf{correlation conflicts} by introducing and integrating \textbf{Follow-up Performance Trends}. Instead of inferring future performance from a decayed or context-enriched representation of the past, FINER takes a novel \textbf{forward-looking} perspective. It directly queries the entire historical dataset to aggregate the empirical outcomes of all students who have exhibited a similar learning pattern.

\section{PRELIMINARIES}
\label{sec:PRELIMINARIES}
A KT setting includes a set of students $S=\{s_1, s_2, \ldots, s_{|S|}\}$ and a set of questions $\mathcal{O}$. Given $o\in \mathcal{O}$ and a binary label $r$, $(o, r)$ is a \textbf{learning cell}, 
where $r = 1$ indicates a correct answer and $r = 0$ indicates a wrong one. $X^s = \langle {x^s_1}, x^s_2, ..., x^s_{|X^s|} \rangle$ represents the \textbf{historical learning sequence} of $s$ in chronological order, where $x^s_k = (o, r)$ is the $k^{\textit{th}}$ learning cell in $X^s$.
$\mathcal{X}=\{X^s|\, s\in S\}$ is the set of historical learning sequences of all students. 
A key KT problem is to predict the probability of students correctly answering a specific question, based solely on pre-target question performance. 
In contrast, we propose to integrate FPTs related to a set of learning patterns of students, extracted from ${X}^s$, which capture the students' post-target question performance in relation to their recent patterns. 

\begin{example}~\label{ex:definition-1}
Continuing Example~\ref{ex:motivation1-2}, Fig.~\ref{fig:sec1-toyexample} shows the historical learning sequence of $ s = \text{Jane}$, i.e.,  ${X}^s=\langle ( o_2, 0 ), ( o_2, 1 ),$ $( o_2,  1 ),  ( o_1, 0 ), ( o_1, 1 )\rangle$.
\end{example}

\begin{definition}  
[\textbf{Learning pattern}] \label{def:learning_pattern}~
A learning pattern $v$ is a sequence of learning cells. 
Given a student $s$ with learning sequence ${X}^s = \langle x^s_1, x^s_2, ..., x^s_R \rangle$ and $i=R-k+1$, then $v_{i}^s=\langle x^s_k, \ldots, x^s_{R} \rangle$ is the $i^{\textit{th}}$ learning pattern of $s$.  $V^s=\{v_i^s| 1 \leq i \leq \min\{\bar{i}, R\}\}$ is the set of learning patterns of $s$, where $\bar{i}$ is a pre-defined parameter specifying the maximum length of a student's learning pattern, and $\mathcal{V}=\{V^s| s\in S  \}$ is the set of learning patterns of a student set $S$.
\end{definition}

Continuing Example~\ref{ex:definition-1} and given $\bar{i}=2$, we have $v_1^s= \langle (o_1,1) \rangle$, $v_2^s= \langle (o_1,0), (o_1,1) \rangle$, and ${V}^s=\{v_1^s, v_2^s\}$. Given $\bar{i}=5$, we have $v_5^s={X}^s$ and ${V}^s=\{v_1^s, v_2^s, v_3^s, v_4^s, v_5^s\}$.

\begin{definition}
[\textbf{Attempt}]~
Given a learning pattern $v$, and a learning sequence ${X}^s$, if $o$ is the $z^{\textit{th}}$ attempt after $v$, then $v \subset {X}^s \wedge (o, r)\in {X}^s \wedge r\in \{0,1\}$ and $(o, r)$ is the $z^{\textit{th}}$ learning cell after $v$.
\end{definition}

Continuing Example~\ref{ex:definition-1} and given a learning pattern $v=\langle (o_2, 1 ), ( o_2, 1 )\rangle$, $o_1$ are the first and second attempts after $v$.

\begin{figure}[!t]
    \centering
    \includegraphics[width=0.47\textwidth]{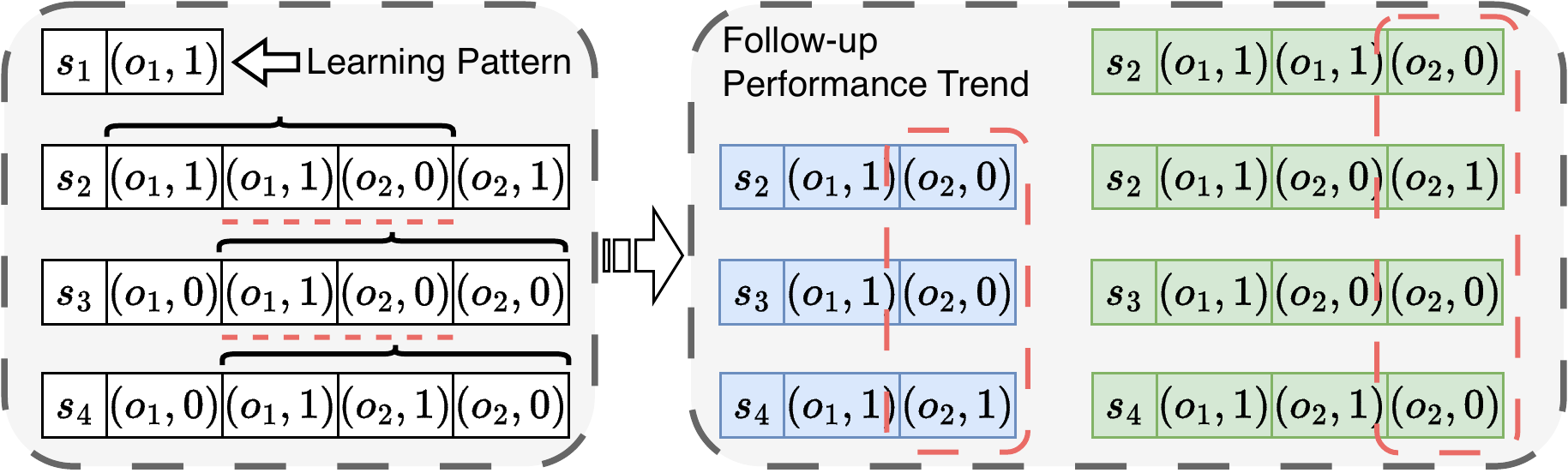}
    \caption{An example of FPT.  Given a student set $S=\{s_2, s_3, s_4\}$, the corresponding historical learning sequence set $\mathcal{X}=\{X^s|\, s\in S\}$, a learning pattern $v=v_1^{s_1}=\langle (o_1, 1) \rangle$, a target question $o_2$, and $\bar{i}=1$ and $\bar{z}=2$, ${T}_{o_2}^{s}=\{t^{v}_{o_2}\}=\{(1, [\frac{1}{3}, \frac{1}{4}], [3, 4])\}$.}
    \label{fig:ExampleForOmega}
\end{figure}

\begin{figure*}[!ht]
    \centering 
    \includegraphics[width=1.0\textwidth]{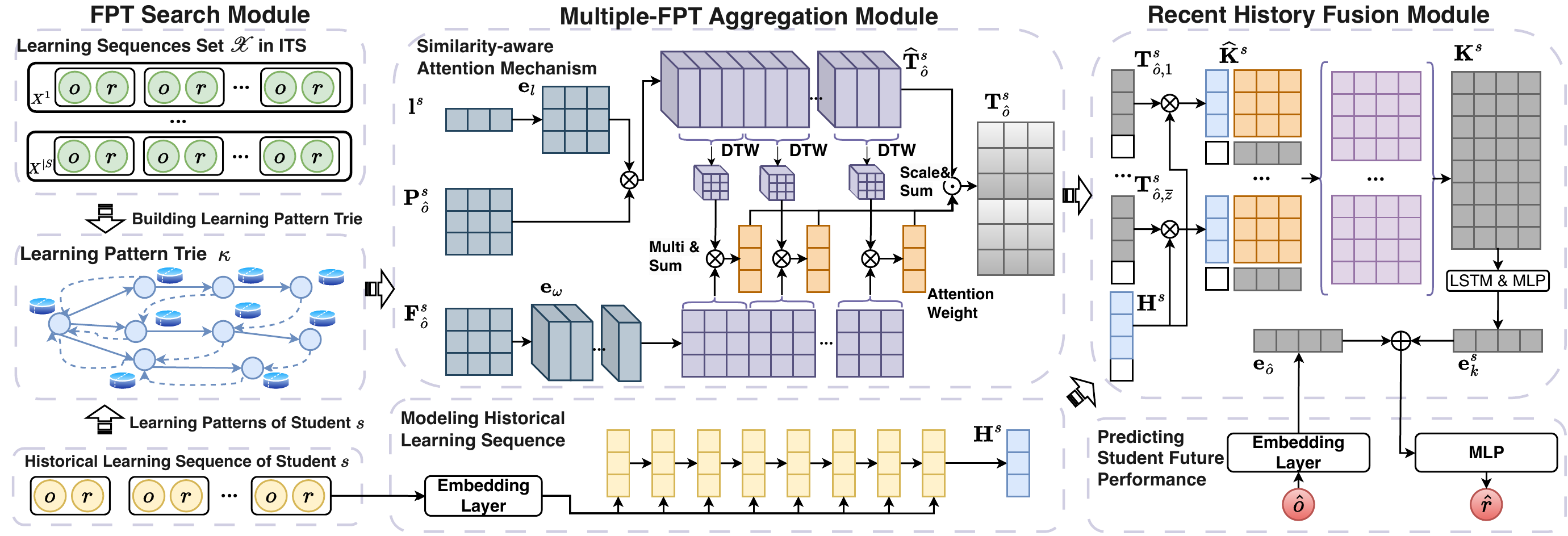}
    \caption{Overview of the \textsf{FINER} Framework.}
    \label{fig:IntroFig1}
\end{figure*}

\begin{definition}
[\textbf{Follow-up Performance Trend}] 
Given a set $\mathcal{X}$ of historical learning sequences of all students,  a learning pattern  \(v\), a target question $\hat{o}$, and a parameter $\bar{z}$, the \textbf{\underline{F}}ollow-up \textbf{\underline{P}}erformance \textbf{\underline{t}}rend (\textbf{FPT}) \(t_{\hat{o}}^{v} = (l^{v}, \omega^{v}_{\hat{o}},\rho^{v}_{\hat{o}})\) records student performance statistics on $\hat{o}$ w.r.t. \(v\). Specifically, 
(i) $l^{v}$ denotes the length of \(v\); 
(ii) vector $\omega^{v}_{\hat{o}}=(\omega^{v}_{\hat{o}}[z]| 1\leq z \leq \bar{z})$, where $\omega^{v}_{\hat{o}}=\omega_{{\hat{o}},0}^{v} + \omega_{\hat{o},1}^{v}$, $\omega_{\hat{o}, 0}^{v}$ denotes the frequency of $ (\hat{o}, 0)$ and $\omega_{\hat{o}, 1}^{v}$ denotes the frequency of $ (\hat{o}, 1)$ such that $\hat{o}$ is the $z^{\textit{th}}$ $(1\leq z \leq \bar{z})$ attempt after \(v\); and 
(iii) vector $\rho^{v}_{\hat{o}}=(\rho^{v}_{\hat{o}}[z]| 1\leq z \leq \bar{z})$, where $\rho^{v}_{\hat{o}}[z] =\omega_{\hat{o}, 1}^{v}[z]/\omega^{v}_{\hat{o}}[z] \,(1\leq z \leq \bar{z})$. 
Given a parameter $\bar{i}$, the set of FPTs for student $s$ on $\hat{o}$ w.r.t. the set of learning patterns 
${V}^s=\{v_i^s|1\leq i\leq \min\{\bar{i}, |{X}^s|\}\}$
is denoted as $T_{\hat{o}}^{s}=\{t_{\hat{o}}^{v}| v\in {V}^s\}$. 
\end{definition}

\begin{example} 
\label{ex:follow_up_performance_trend} Fig.~\ref{fig:ExampleForOmega} exemplifies a set of historical learning sequences  $\mathcal{X}=\{ X^{s_2}, X^{s_3}, X^{s_4}\}$ of three students $s_2$, $s_3$, and $s_4$, and a historical learning sequence $X^{s_1}$ of student $s_1$. 
Given $\bar{i} = 1$ and a learning pattern $v=v_1^{s_1}=\langle (o_1, 1) \rangle$, a target question $o_2$, and $\bar{z}=2$, then $t^{v}_{o_2}=(l^{v}, \omega^{v}_{o_2},\rho^{v}_{o_2})$.
Specifically, 
(i) $l^{v}=1$ as there is one learning cell in $v$; (ii) $\omega^{v}_{o_2}= [3, 4]$ as $\langle(o_1, 1), (o_2, r)\rangle\, (r \in \{0,1\})$ occurs three times, i.e., $\langle(o_1, 1), (o_2, 0)\rangle $, $\langle(o_1, 1), (o_2, 0)\rangle$, and $\langle(o_1, 1),$ $(o_2, 1)\rangle$, and $\langle(o_1, 1), (o, r), (o_2, r)\rangle\, (o\in \mathcal{O}, r \in \{0,1\})$ occurs four times, i.e., $\langle(o_1, 1), (o_1, 1), (o_2, 0)\rangle$,  $\langle(o_1, 1), (o_2, 0), (o_2, 1)\rangle$, $\langle(o_1, 1), (o_2, 0), (o_2, 0)\rangle$, and $\langle(o_1, 1), (o_2, 1), (o_2, 0)\rangle$; and 
(iii) $\rho^{v}_{o_2}=[\frac{1}{3}, \frac{1}{4}]$ due to $\omega_{o_2, 1}^{v}[1]=\omega_{o_2, 1}^{v}[2]=1$. ${T}^{s_1}_{o_2}=\{t^{v}_{o_2} | v\in {V}^{s_1}\}$, where ${V}^{s_1}=\{v_1^{s_1}\}=\{\langle (o_1, 1) \rangle\}$. 
\end{example}

\vspace{0.5em}
\begin{definition}[\textbf{Forward-Looking Knowledge Tracing}]~\label{def2}~
Given a set  $\mathcal{X}$ of historical learning sequence, a historical learning sequence ${X}^s$ of a student $s$, a target question $\hat{o}\in \mathcal{O}$, and parameters $\bar{z}$ and $\bar{i}$,  the
 \textbf{\underline{F}}orward-Look\textbf{\underline{i}}ng K\textbf{\underline{n}}owledg\textbf{\underline{e}} T\textbf{\underline{r}}acing (\textsf{FINER}) framework: 

\begin{itemize}[itemsep=1pt, leftmargin=12pt]
\item finds the set of  learning patterns ${V}^s=\{v_i^s| 1 \leq i \leq \bar{i}\}$ of $s$, and obtains the set of FPTs ${T}^s_{\hat{o}}=\{t_{\hat{o}}^{\smash{v_i^s}}|1\leq i\leq\bar{i} \}$ from $\mathcal{X}$. 
\item predicts the probability $\alpha_{\hat{o}}^{s}$ based on ${T}_{\hat{o}}^s$ and ${X}^s$, where $\alpha_{\hat{o}}^{s}$ is the probability that $s$ answers $\hat{o}$ correctly after ${X}^{s}$.
\end{itemize}
\end{definition}

\begin{example} 
Continuing Example~\ref{ex:follow_up_performance_trend}, \textsf{FINER} first finds ${V}^{s_1}$ and obtains ${T}_{o_2}^{s_1}$. Following this, \textsf{FINER} predicts $\alpha_{{o_2}}^{s_1}$ based on ${T}_{o_2}^{s_1}$ and ${X}^{s_1}$. If $\alpha_{{o_2}}^{{s_1}}=0.5$, this implies that $s_1$ has a 50\% probability of correctly answering ${o_2}$ after experiencing ${X}^{s_1}$.
\end{example}
\section{Forward-Looking Knowledge Tracing}
\label{sec:FINER}
This section presents \textsf{FINER} that integrates FPTs with historical learning sequences to enable more accurate student performance prediction. The key innovation is a three-module architecture that efficiently retrieves, aggregates, and fuses FPTs with historical data. The modules are detailed in Sections~\ref{sec:FPT Search Module}, \ref{sec:Multiple-FPT Aggregation Module}, and \ref{sec:FPT and Individual History Fusion Module}.

\begin{figure}[!tbp]
    \includegraphics[width=0.44\textwidth]{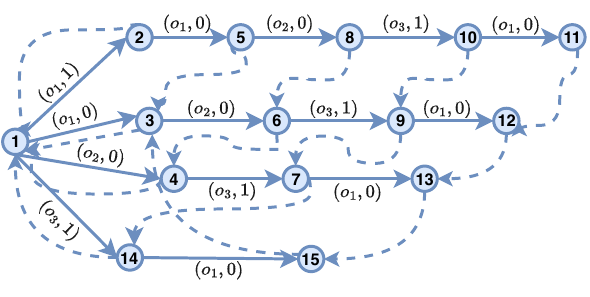}
    \label{fig:Building Learning Pattern Automaton}
    \caption{A learning pattern trie built based on Fig.~\ref{fig:sec1-toyexample}. (i) Circles represent nodes. (ii) Arrows represent edges and each dashed line represents a virtual edge $a\rightarrow \tilde{a}$ that is not stored in the trie
 (iii) $\bar{i} = 2$ and $\bar{z} = 3$.} 
    \label{fig:ExampleOfLearningPatternAutomaton} 
\end{figure}

\subsection{FPT Search Module} \label{sec:FPT Search Module}
The aim of the FPT Search Module (left side of Fig.~\ref{fig:IntroFig1}) is to efficiently retrieve a student's FPTs from a set of learning sequences $\mathcal{X}$. We design a learning pattern trie to compress and save key information from these sequences and propose a novel algorithm for real-time FPT retrieval.

\subsubsection{Learning Pattern Trie Construction} \label{sec:Learning Pattern Trie Construction}
\label{sec: Building Learning Pattern Automaton} 
Given a set of historical learning sequences $\mathcal{X}$, a learning pattern $v$, and the parameter $\bar{z}$, retrieving the FPT $t^v$ of $v$ from $X^s\in \mathcal{X}$ essentially corresponds to matching $v$ with subsequences in $X^s$ and then recording the subsequent $\bar{z}$ learning cells after each matched subsequence. The time complexity of a naive approach to performing this procedure is \(O (\sum_{s\in S}(|X^s| + l^v + \bar{z})) \).

\begin{definition}
[\textbf{Learning pattern trie}] \label{def:trie}
Given a set of questions $\mathcal{O}$, a set of historical learning sequences $\mathcal{X}$, a set of students $S$, and a parameter $\bar{z}$, 
a learning pattern trie 
is given by $\kappa = (\mathcal{A}, \mathcal{B})$, where $\mathcal{A}$ is a set of nodes and $\mathcal{B}$ is a set of edges. 
Each edge $b=(a\rightarrow a') \in \mathcal{B}$ is a learning cell $(o,r)$.  $a_r\in \mathcal{A}$ is the root node of $\kappa$. Each node $a\in {\mathcal{A}/\{a_r\}}$ is associated with a unique learning pattern $v=\langle(a_r\rightarrow a'), (a'\rightarrow a''), \ldots, (a'''\rightarrow a)\rangle\,(v\subset X^s \wedge s\in S)$ and stores the node information $(t^{v},  \eta^{v},  \tilde{a})$.
\begin{itemize}[itemsep=1pt, leftmargin=12pt]
\item  $t^v=(l^{v}, \omega^{v}, \rho^{v})$ is the FPT of all questions $o\in \mathcal{O}$ w.r.t. $v$. Specifically,  (i) $l^{v}$ is the length of $v$; (ii) $\omega^{v}= [(\omega^{v}_{o_m})^{\rm{T}}| o_m\in \mathcal{O}, 1\leq m\leq |\mathcal{O}|]$, 
 where  $(\omega^{v}_{o_m})^{\rm{T}}$ is the $m^{\textit{th}}$ column of $\omega^{v}$; and (iii)  $\rho^{v} = [(\rho^{v}_{o_m})^{\rm{T}}| o_m \in \mathcal{O}, 1\leq m\leq |\mathcal{O}|]$, 
 where  $(\rho^{v}_{o_m})^{\rm{T}}$ is the $m^{\textit{th}}$ column of $\rho^{v}$ such that $\rho^{v}_{{o_m}}[z] =\omega_{{o_m}, 1}^{v}[z]/\omega^{v}_{{o_m}}[z] \,(1\leq z \leq \bar{z})$.
\item $\eta^{v}$ is the frequency of $v$ in $\mathcal{X}$. 
\item $\tilde{a}\, (\tilde{a}\in \mathcal{A})$ is defined as a suffix node of $a$, such that the pattern $\tilde{v}\,(\tilde{v}\in \mathcal{V})$ associated with $\tilde{a}$ is the suffix of $v$ with a length of $l^v-1$, i.e., $\tilde{v}=\langle(a_r\rightarrow a'), (a'\rightarrow a'')\ldots, (a'''\rightarrow a)\rangle$. Finally, $\tilde{a}=a_r$ if $l^v = 1 \vee a=a_r$.
\end{itemize}
\end{definition}

\begin{example}~\label{exp:ToyExampleLearningPatternAutomaton}
Given $\mathcal{O}=\{o_1, o_2, o_3\}$, $X= \{\langle( o_1, 1 ), ( o_1, 0 ), ( o_2, 0),$ $( o_3,  1 ),  ( o_1, 0 )\rangle\}$, and $\bar{z}=3$, 
Fig.~\ref{fig:ExampleOfLearningPatternAutomaton} 
shows the learning pattern trie. Node $5$ is associated with the learning pattern $v=\langle (o_1, 1), (o_1, 0) \rangle$ and stores the node information $(t^v, \eta^{v}, \tilde{a})$, where $t^v=(l^{v}, \omega^{v}, \rho^{v})$. 
Specifically, (i) $l^v=2$; (ii)  $\omega^{v}= [(\omega^{v}_{o_1})^{\rm{T}}, (\omega^{v}_{o_2})^{\rm{T}}, (\omega^{v}_{o_3})^{\rm{T}}]$, where $\omega^{v}_{o_1}=[0,0,1]$, $\omega^{v}_{o_2}=[1,0,0]$, and $\omega^{v}_{o_3}=[0,1,0]$; (iii) $\rho^{v}= [(\rho^{v}_{o_3})^{\rm{T}}, (\rho^{v}_{o_2})^{\rm{T}}, (\rho^{v}_{o_3})^{\rm{T}}]$, where $\rho^{v}_{o_1}=[0,0,0]$, $\rho^{v}_{o_2}=[0,0,0]$, and $\rho^{v}_{o_3}=[0,1,0]$; (iv) $\eta^{v} = 1$; and (v) $\tilde{a}=3$ as the suffix of $v$ of length 1 is represented by node 3. Note that each dashed line represents a virtual
 edge $a\rightarrow \tilde{a}$ that is not stored in the trie.
\end{example}
  
Algorithm \ref{alg:BuildingLearningPatternExtractionAutomata} describes the construction of a learning pattern trie from a set  $\mathcal{X}$ of historical sequences {(see the middle-left of Fig.~\ref{fig:IntroFig1})}.
First, $\kappa=\{\mathcal{A}, \mathcal{B}\}$ is initialized, where $\mathcal{A}$ contains a root node $a_r$, and $\mathcal{B}$ is an empty set (line 1). Second, the algorithm examines each learning cell $x_k^s$ in each historical learning sequence ${X}^s\in \mathcal{X}$ (lines 2--13) to identify and add learning patterns to the trie. The adjacency information for a node $a$ is maintained in a hash table. Given the current learning cell $x_k^s=(o_m, r)$, its key is computed as $(r\times |\mathcal{O}| + m)\bmod \textit{hz}$, where $\textit{hz}$ is the dynamically adjusted size of the hash table based on the node set size of the trie. In line 11,
$v +  x^s_k$ indicates that the current learning cell $x_k^s$ is appended to the current learning pattern $v$. In particular, the learning pattern associated with each node has at most $\xi$ consecutive identical learning cells, i.e., 
$\forall a \in \mathcal{A} ( \nexists a' \in \mathcal{A} ((v=\langle (a'\rightarrow a''), \ldots, (a'''\rightarrow a)\rangle \wedge l^v \geq \xi) \Rightarrow  (a'\rightarrow a'')=\ldots= (a'''\rightarrow a)))$ (lines 6--7). Next, the algorithm applies the Failure function $f(\cdot)$~\cite{DBLP:journals/cacm/AhoC75} to compute the suffix node $\tilde{a}$ for $a \in \mathcal{A}$ (lines 14--15). Finally, $t^v$ of each pattern $v$ associated with $a\in{\mathcal{A}}$ is computed using a function $\texttt{g}(\cdot)$ (line 16).

\setlength{\textfloatsep}{-1pt} 
\setlength{\intextsep}{-1pt} 
\setlength{\floatsep}{-1pt} 
\begin{algorithm}[!t]
  \caption{Constructing A Learning Pattern Trie}
  \label{alg:BuildingLearningPatternExtractionAutomata}
  \KwInput{The set of historical learning patterns $\mathcal{X}$ and parameters $\xi$ and $\bar{z}$. }
  \KwOutput{A learning pattern trie $\kappa=(\mathcal{A}, \mathcal{B})$.}
  $\mathcal{A} = \{a_r\}$, $\mathcal{B} = \emptyset$; 
  \bluecomment{Initialize $\kappa$}\\
  \ForEach{$X^s \in \mathcal{X}$}{
      $a \leftarrow a_r$, $v = \langle \rangle$; \\
      \For {${k} = 1$ \KwTo $|X^s|$}{  
          {
           \uIf{$k\geq \xi$}{
              \uIf{$x_k^s=x_{k-1}^s=\ldots=x_{k-\xi+1}^s$}{
                  continue\;
              }
           }   
              \uIf {$\nexists a' \in \mathcal{A} ((a \rightarrow a') = x^s_k)$}{
                  $\mathcal{A} \leftarrow \mathcal{A} \cup \{ a' \}$; $\mathcal{B} \leftarrow \mathcal{B} \cup \{ (a \rightarrow a') \}$\;
                  
                  $v' \leftarrow v +  x^s_k $, $\eta^{v'} \leftarrow 0$\;
              }
              $\eta^{v} \leftarrow \eta^{v} + 1$; $a \leftarrow a'$; $v \leftarrow v +  x^s_k $; \\
              $\omega^{v}_{o, r}[z] = 0 \, (o \in \mathcal{O} \wedge r \in \{0,1\} \wedge 1\leq z\leq \bar{z})$\; 
             $\rho^{v}_{o, r}[z] = 0 \, (o \in \mathcal{O} \wedge r \in \{0,1\} \wedge 1\leq z\leq \bar{z})$\;
  
          }
      }
  }
  \ForEach{$a \in \mathcal{A}$}{  $\tilde{a}\leftarrow f(a)$; \bluecomment{Find the suffix node}\\
  }
  \texttt{g} ($a_r$, $\bar{z}$, $\langle \rangle$, 0)\;
  \KwRet $\kappa$\;
  \end{algorithm}
  
  \begin{algorithm}[!t]
  \caption{{Computing FPTs}}
  \label{alg:ObtainingFuturePerformanceTrends}
  \KwInput{The node $a$, a parameter $\bar{z}$, the learning pattern $v$ associated with $a$, and the length $l$ of $v$.}
  \KwOutput{The FPT $t^v$ of learning pattern $v$.}
  \SetKwFunction{Fg}{g}
  \SetKwProg{Fn}{Function}{:}{}
  \Fn{\Fg{$a$, $\bar{z}$, $v$, $l$}}{
    \ForEach{$(a \rightarrow a') = (o,r) \in \mathcal{B}$}{
      {$v' = {v} + \langle (o,r) \rangle$,\, $\omega_{o,r}^{v}[1] = \eta^{v'}$;\\
      {
      $\tau_{a}[1] = \tau_{a}[1] \cup \{(o, r)\}$\;
      }
      \Fg($a', \bar{z}, v', {l + 1})$}\;
      \For{$z = 2$ \KwTo $\bar{z}$}{
        {$\tau_{a}[z] = \tau_{a}[z] \cup \tau_{a'}[z-1]$}\;
        \ForEach{$({o}', {r}')$ in $\tau_{a'}[z-1]$}{
          $\omega^{v}_{{o}', {r}'}[z] = \omega^{v}_{{o}', {r}'}[z] + \omega^{v'}_{{o}', {r}'}[z-1]$\;
        }
      }
    }
    {$l^v = l$};\, \bluecomment{The length of $v$} \\
    \ForEach{$o \in \mathcal{O}$}{
      $\rho_{o}^{v} = \omega_{o, 1}^{v}/ \omega_{o}^{v}$\;
    }
    \KwRet $t^v$; 
  }
  \end{algorithm}
\setlength{\textfloatsep}{0pt} 
\setlength{\intextsep}{0pt} 
\setlength{\floatsep}{0pt} 
Algorithm~\ref{alg:ObtainingFuturePerformanceTrends} details the procedure of computing FPTs using the recursive function $\texttt{g}(\cdot)$. It initiates from the root node $a_r$ of the learning pattern trie, a learning pattern $v=\langle \rangle$ (line 16, Algorithm~\ref{alg:BuildingLearningPatternExtractionAutomata}). Next, since $a'$ is the $1^\textit{st}$ order child of $a$ with $(a\rightarrow a')=(o,r)$, $\omega^v_{o,r}[1]$ is set to $\eta^{v'}$ and $(o,r)$ is added to $\tau_{a}[1]$, where $v$ and $v'$ are the learning patterns associated with $a$ and $a'$, respectively (lines 3--4). After $\texttt{g}(\cdot)$ recursively calls itself, \(\tau_{a}[z]\) and $\omega^{v}_{{o}', {r}'}[z]$ are updated using 
\(\tau_{a}[z']\, (1 \leq z' \leq \bar{z}-1)\)  and $\omega^{v'}_{{o}', {r}'}[z-1]\, \, (2 \leq z \leq \bar{z})$, respectively (lines 5--8). Finally, \(\rho\) is derived according to $ \omega_{o, 1}^{v}$ and $ \omega_{o}^{v}$ (line 9-10). The overall time complexity of Algorithm~\ref{alg:ObtainingFuturePerformanceTrends} is $O(\sum_{a \in  \mathcal{A}}(|\mathcal{O}| + \sum_{1 \leq z \leq \bar{z}}|\tau_{a}[z]|)) $ and thus the  overall time complexity of Algorithm~\ref{alg:BuildingLearningPatternExtractionAutomata} is {$O(\sum_{s\in S}|X^s| + \sum_{a \in  \mathcal{A}}(|\mathcal{O}| + \sum_{1 \leq z \leq \bar{z}}|\tau_{a}[z]|))$}.

\subsubsection{Fetching FPTs in Real Time} 
\label{subsec:FuturePerformanceExtractor}
Given a learning pattern trie $\kappa$, a newly arrived learning cell $x^s_k$, and a historical learning sequence ${X}^s$ of a student $s$, a target question $\hat{o}$, parameters $\xi$, $\bar{z}$, and $\bar{i}$, the module aims to fetch the FPTs of learning patterns {$V^s = \{v^s_i | 1 \leq i \leq \bar{i} \}$ w.r.t. $\hat{o}$ for each updated learning sequence ${X}^s =\langle x_1^s, \ldots, x_k^s \rangle$ in real time using $\kappa$ (see the bottom-left of Fig.~\ref{fig:IntroFig1}).  

\setlength{\textfloatsep}{-1pt} 
\setlength{\intextsep}{-1pt} 
\setlength{\floatsep}{-1pt} 
\begin{algorithm}[!t]
  \caption{Fetching FPTs in Real-time}
  \label{alg:PotentialFuturePerformanceExtractor}
  \KwInput{A learning pattern trie $\kappa$, an arrived learning cell $x^s_k$, a historical learning sequence $X^s$ of a student $s$, 
  a target question $\hat{o}$, parameters $\xi$, 
  $\bar{z}$, and $\bar{i}$, and a node $a$. 
  }
  \KwOutput{The FPT set $T^s_{\hat{o}}$.}
  ${X}^s \leftarrow {X}^s + x_k^s$; \bluecomment{Update ${X}^s$}\; 
  {
      \uIf{$k\geq \xi$}{
          \uIf{$x_k^s=x_{k-1}^s=\ldots=x_{k-\xi+1}^s$}{
              \KwRet $(T^{s}_{\hat{o}})$\;
          }
      }
      \While{$(\nexists a'\in \mathcal{A} ((a \rightarrow a') = x^s_k) \vee l^{v'} > \bar{i}) \wedge a \neq a_r$}{
             $a \leftarrow \tilde{a}$;  \bluecomment{Trace the suffix node}
          }
          $a'' \leftarrow a'$; $a \leftarrow a'$; 
       
          \For{$i = 1$ \KwTo $\bar{i}$}{
             $T_{\hat{o}}^{s} \leftarrow T_{\hat{o}}^{s} \cup \{t_{\hat{o}}^{v''}\}$; $a'' \leftarrow \tilde{a}''$\;
          }
       
          {$a\leftarrow a'$;  \bluecomment{Maintain $a$}\\
          \KwRet $T^{s}_{\hat{o}}$; 
          }
  }
\end{algorithm}
\setlength{\textfloatsep}{0pt} 
\setlength{\intextsep}{0pt} 
\setlength{\floatsep}{0pt} 

Algorithm~\ref{alg:PotentialFuturePerformanceExtractor} details the process. Initially, node $a$ corresponds to a pattern $v$, which is the longest suffix of ${X}^s$ found in $\kappa$ with a length $l^v \leq \bar{i}$.
When a learning cell $x^s_k$  arrives, the learning pattern and its FPT are not updated if the most recent $\xi$ learning cells are the same (lines 3--4). 
This is in line with the principle of constructing $\kappa$, where each learning pattern contains at most $\xi$ learning cells. 
Next, the algorithm identifies the node $a'$ in $\kappa$ representing $v'$; 
if $\nexists a'\in \mathcal{A}\,( (a \rightarrow a') = x^s_k$), it checks the suffix with length $l^v-1$, corresponding to $\tilde{a}$ (lines 5--6).  The idea is to always match a longer suffix of $v$ with a pattern in $\kappa$. 
Finally, we {get} the FPT set ${T}^s_{\hat{o}}=\{t_{\hat{o}}^{v_i^s}|1\leq i\leq\bar{i} \}$ (lines 8--11). 
We trace the suffixes of $v_i^s$ via suffix nodes $\tilde{a}$. Thus, if ${a}''=a_r$ or $i\leq \bar{i}$ and $a$ represents $v^s_{i}$, this implies that $v_i^s\,( 0\leq i \leq \bar{i})$ does not exist in $\kappa$; in this case, the algorithm loops at $a_r$ for $\bar{i}-i$ times and sets $v_i^s=v'\,( 0\leq i < i')=\langle \rangle$, where $\langle \rangle$ is the pattern corresponding to $a_r$ (see Section~\ref{sec: Building Learning Pattern Automaton}). Note that this scenario is rare, as the size of the historical learning sequence set $\mathcal{X}$ is generally very large. 

\noindent\textbf{{Time complexity.}} Using a learning pattern trie improves the efficiency of fetching FPTs. Specifically, without the trie, traversing each historical learning sequence $X^s$ from $\mathcal{X}$ to find the longest suffix $v$ ($l^v \leq \bar{i}$) of ${X}^s$ yields a time complexity of $O(\sum_{s \in S} (|{X}^s| +\bar{i} + \bar{z}))$. In contrast, Algorithm~\ref{alg:PotentialFuturePerformanceExtractor} achieves a time complexity of $O(\bar{i})$. This is because the trie allows both retrieving the longest suffix $v$ ($l^v \leq \bar{i}$) of ${X}^s$ and fetching the $\bar{z}$ attempts after $v$ in just $O(\bar{i})$ time each.
\subsection{Multiple-FPT Aggregation Module} \label{sec:Multiple-FPT Aggregation Module}
\label{subsec:TrendAggregateModule}
The Multiple-FPT Aggregation Module (in the middle of Fig.~\ref{fig:IntroFig1}) first embeds the FPTs of $\bar{i}$ learning patterns returned by the search module and then aggregates the FPTs according to their confidence scores. To assess the confidence, we design a similarity-aware attention mechanism.

\subsubsection{Representation Learning of FPTs}
\label{sec:RepresentationOfFPTs}
To facilitate the subsequent combination of FPTs with  learning sequences, the aggregation module initially embeds the FPTs as a distributed representation.
We start by embedding  \(\mathbf{l}^s\)  to learn the representation of FPTs. Since \(\mathbf{l}^s\) is a discrete variable, we one-hot encode it as $\mathbf{e}_l              = \texttt{one-hot}(\mathbf{l}^s) \cdot \mathbf{W}_{l}$, 
where \(\mathbf{e}_l \in \mathbb{R}^{\bar{i} \times d}\) is the embedding of \(\mathbf{l}^s\), $d$ is a hyper-parameter, $\texttt{one-hot}(\cdot)$ denotes one-hot encoding, and $\cdot$ is matrix multiplication. Next, $\mathbf{e}_l$ is combined with \(\mathbf{P}^s_{\hat{o}} \in \mathbb{R}^{\bar{i} \times \bar{z}}\) through matrix multiplication to capture the dynamic changes of different FPTs over attempts, where $\mathbf{P}^s_{\hat{o},i}$ denotes the $i^{\textit{th}}$ row of \(\mathbf{P}^s_{\hat{o}}\). To align with the operation requirements,  we reshape $\mathbf{P}^s_{\hat{o}}$ into \(\mathbb{R}^{\bar{i} \times \bar{z} \times 1}\) and \(\mathbf{e}_l\) into \(\mathbb{R}^{\bar{i} \times 1 \times d}\). The reshaped tensors are then ready for the subsequent matrix multiplication, i.e., $\hat{\mathbf{T}}^s_{\hat{o}} = \mathbf{P}^s_{\hat{o}} \cdot \mathbf{e}_l$, 
\noindent where $\hat{\mathbf{T}}^s_{\hat{o}} \in \mathbb{R}^{\bar{i} \times \bar{z} \times d}$  represents the embedding of FPTs. 

\subsubsection{Similarity-aware Attention Mechanism}\label{sec:Similarity-aware_Attention_Mechanism}

The frequency of each FPT is captured by its corresponding element in $\mathbf{F}^s_{\hat{o}}$. 
However, simply using the frequency of an FPT as its confidence score can lead to underestimating less frequent but relevant ones. Hence, it is inappropriate to use these frequencies as the sole indicators of confidence. 

Thus, instead of relying solely on frequency, we also consider the relationships between FPTs. Specifically, FPTs that are similar to each other tend to receive similar confidence scores, even if some of them may be less frequent. Thus, we employ DTW~\cite{DBLP:journals/talg/GoldS18} to assess the similarity between adjacent trends $v^s_i$ and $v^s_{i'}$ $(v^s_i, v^s_{i'}\in {V}^s \wedge |i-i'|\leq \lambda)$ and to form a similarity matrix \( D^s_{\hat{o},i,i'}\), where $\lambda$ is a parameter defining the adjacency of trends {(see the upper-middle of Fig.~\ref{fig:IntroFig1})}: 

\begin{equation}
\begin{aligned}
&\mathbf{D}^s_{\hat{o},i,{i'}}(z, z') = {cos}(\hat{\mathbf{T}}^s_{\hat{o},i}(z), \hat{\mathbf{T}}^s_{\hat{o},i'}(z')) + \\
& \max(\mathbf{D}^s_{\hat{o},i,i'}(z-1, z'-1), \mathbf{D}^s_{\hat{o},i,i'}(z, z'-1), \mathbf{D}^s_{\hat{o},i,i'}(z-1, z')) \\
\end{aligned}
\end{equation}
Here, $\hat{\mathbf{T}}^s_{\hat{o},i} \in \mathbb{R}^{\bar{z} \times d}$ is the representation of $t^{v^s_i}$, where $d$ is a hyperparameter. Next, $\hat{\mathbf{T}}^s_{\hat{o},i}(z) \in \mathbb{R}^{d}$ is the $z^{\textit{th}}$ column of $\hat{\mathbf{T}}^s_{\hat{o},i}$, which denotes the representation of $t^{v^s_i}$ on $\hat{o}$, such that $\hat{o}$ is the $z^{\textit{th}}$ attempt after $v^s_i$. 
Further, $\hat{\mathbf{T}}^s_{\hat{o},i}(1:z) \in \mathbb{R}^{z \times d}$ is the first $z$ columns of $\hat{\mathbf{T}}^s_{\hat{o},i}$, and
${cos}(\hat{\mathbf{T}}^s_{\hat{o},i}(z), \hat{\mathbf{T}}^s_{\hat{o},i'}(z'))$ represents the cosine similarity between $\hat{\mathbf{T}}^s_{\hat{o},i}(z)$ and $\hat{\mathbf{T}}^s_{\hat{o},i'}(z')$.  
Finally, $\mathbf{D}^s_{\hat{o},i, i'}(z, z') \in \mathbb{R}^{\bar{z} \times \bar{z}}$ is the similarity between $\hat{\mathbf{T}}^s_{\hat{o},i}(1:z)$ and $\hat{\mathbf{T}}^s_{\hat{o},i'}(1:z') \,(1 \leq z, z' \leq \bar{z})$. 

Next, given the high frequency of certain FPTs leading to large values in $\mathbf{F}^s_{\hat{o}}$, direct embedding is inappropriate. Instead, we logarithmically scale $\mathbf{F}^s_{\hat{o}}$, round it down, and then embed it as follows.  

\begin{equation}\label{fun:confidence_level}
\begin{aligned}
\mathbf{e}_{\omega} & = {\texttt{sigmoid}}(\texttt{MLP}(\mathbf{W}_{\mathbf{\omega}} \lfloor \log_2 ({\mathbf{F}_{\hat{o}}^s} \cdot 10 \rfloor))),\\
\end{aligned}
\end{equation}
where  $\texttt{sigmoid}(\cdot)$ is the sigmoid activation function and $\mathbf{e}_{\omega} \in \mathbb{R}^{\bar{i} \times \bar{z}}$ is the 
embedding frequency of the FPTs. We compute the attention weight of an FPT $t^{v^i_s}$ based on the similarity matrix: 
\begin{equation}
\begin{aligned}
{\textit{att}^{s}_{\hat{o},i }} = \frac{1}{2 \lambda + 1}\sum_{i' = i-\lambda}^{i+\lambda} \mathbf{e}_{\omega,i'} \cdot \mathbf{D}^s_{\hat{o},i,i'}, \\
\label{equ:att}
\end{aligned}
\end{equation}

\noindent where $\mathbf{e}_{\omega,i'}$ is the embedding of $\omega^{v_{i'}^s}_{\hat{o}}$ and
${\textit{att}^{s}_{\hat{o},i }} \in \mathbb{R}^{\bar{z}}$ represents the attention weight of an FPT on  $\hat{o}$ w.r.t. ${v^i_s}$. The similarity-aware estimation ensures that the attention weight for each FPT is determined not only by its frequency but also by its contextual relevance, which is derived from its similarity to adjacent FPTs.

\subsubsection{FPTs aggregation}
After obtaining the distributed representation $\hat{\mathbf{T}}^s_{\hat{o}}$ and attention weights for each FPT, we aggregate them to form a comprehensive representation that captures the overall performance trends. The aggregation process takes into account both the semantic meaning of each FPT (encoded in $\hat{\mathbf{T}}^s_{\hat{o}}$) and its relative importance (determined by the attention weights). Specifically, we compute a weighted sum of the FPT representations, where the weights are derived from the attention mechanism. This weighted aggregation ensures that FPTs with higher confidence scores (based on both frequency and similarity) contribute more significantly to the final representation. The aggregation is performed as follows:

\begin{equation}
\begin{aligned}
\mathbf{T}^s_{\hat{o}} &= \frac{1}{\bar{i}} \sum_{i=1}^{\bar{i}} {\textit{att}^{s}_{\hat{o},i }} \cdot \hat{\mathbf{T}}^s_{\hat{o},i}, \\
\end{aligned}
\end{equation}
where $\mathbf{T}^s_{\hat{o}} \in \mathbb{R}^{\bar{z} \times d}$ represents the final aggregated FPT representation. This representation maintains the temporal structure of the performance trends (through the $\bar{z}$ dimension) while capturing the semantic features (through the $d$ dimension). The normalization factor $\frac{1}{\bar{i}}$ ensures that the aggregation is not biased by the number of FPTs being combined.

The resulting $\mathbf{T}^s_{\hat{o}}$ serves as a comprehensive summary of the student's potential future performance trends, incorporating both the semantic meaning of each trend and its relative importance based on frequency and similarity to other trends.

\subsection{Recent History Fusion Module} \label{sec:FPT and Individual History Fusion Module}
The Recent History Fusion Module (to the right in Fig.~\ref{fig:IntroFig1}) addresses temporal misalignment of FPTs and learning sequences through three key steps: (i) independently encoding learning sequences and FPTs, (ii) fusing their representations via tensor outer products, and (iii) modeling temporal dependencies at the feature level using LSTM networks to predict student performance.

\subsubsection{Historical Learning Sequence Modeling}
After obtaining FPT representations (see Section~\ref{sec:Multiple-FPT Aggregation Module}), we encode students' historical learning sequences into distributed representations to capture their temporal learning dynamics \cite{DBLP:conf/nips/PiechBHGSGS15}.

To model the sequential dependencies and temporal patterns within the learning trajectory, we employ an LSTM network that processes the sequence of embedded learning cells. The LSTM architecture is particularly well-suited for this task as it can effectively capture long-term dependencies while mitigating the vanishing gradient problem commonly encountered in sequential modeling. The embedding of a student's complete historical learning sequence $\mathbf{H}^s \in \mathbb{R}^{d}$ is derived as follows {(see the lower-middle of Fig.~\ref{fig:IntroFig1})}:

\begin{equation}
    \mathbf{H}^s = \texttt{LSTM}([\mathbf{e}_{x_1^s}, \mathbf{e}_{x_2^s},...,\mathbf{e}_{x_{|{X}^s|}^s}]),
\end{equation}
where $|{X}^s|$ represents the total number of learning interactions in student $s$'s historical sequence, and $\texttt{LSTM}(\cdot)$ denotes the LSTM layer that processes the chronologically ordered sequence of embedded learning cells. The resulting representation $\mathbf{H}^s$ encapsulates the student's cumulative learning experience, capturing both the content knowledge acquired and the temporal patterns of their learning behavior.

\subsubsection{FPT and Individual History Fusion}
The fusion of historical learning sequences and FPTs presents a fundamental challenge: how to effectively combine two distinct types of temporal information that operate at different granularities. Historical learning sequences capture individual learning interactions, while FPTs represent aggregated performance patterns over multiple attempts. To address this challenge, we employ a tensor-based fusion mechanism that preserves the semantic richness of both representations while enabling their effective integration.

Inspired by multimodal fusion techniques \cite{zadeh2017tensor}, we utilize tensor outer products to capture complex interactions between historical learning behaviors $\mathbf{H}^s$ and FPT representations $\mathbf{T}^s_{\hat{o}, z}$. The tensor outer product operation naturally models all pairwise interactions between features from both modalities, creating a comprehensive representation that captures both individual feature contributions and their cross-modal dependencies.

Specifically, we augment both representations with bias terms and compute their tensor outer product:
\begin{equation}
\mathbf{\hat{K}}_{z} =
\left [ \mathbf{H}^s, 1 \right ]^T \otimes \left [ \mathbf{T}^s_{\hat{o}, z}, 1 \right ]^T,
\end{equation}
where $\hat{\mathbf{K}}_{z} \in \mathbb{R}^{(d+1) \times (d+1)}$ represents the tensor fusion result that encapsulates the interaction between the historical learning sequence $\mathbf{H}^s$ and the FPT representation $\mathbf{T}^s_{\hat{o}, z}$ at the $z^{\textit{th}}$ attempt. The augmentation with bias terms (represented by the constant 1) enables the model to learn both multiplicative and additive interactions between the two modalities.

The resulting tensor $\hat{\mathbf{K}}_{z}$ contains rich interaction information but requires dimensionality reduction to extract meaningful features for subsequent processing. We apply a learnable transformation to project this high-dimensional tensor into a more compact representation:
\begin{equation}
\begin{aligned}
\mathbf{K}_{z} &= \texttt{ReLU}(\mathbf{W}_{k} \cdot \mathbf{\hat{K}}_{z} + \mathbf{b}_k), \\
\end{aligned}
\end{equation}
where $\texttt{ReLU}(\cdot)$ is the rectified linear unit activation function that introduces non-linearity and sparsity to the fused representation. The parameters $\mathbf{W}_{k} \in \mathbb{R}^{d \times (d+1) \times (d+1)}$ and $\mathbf{b}_{k} \in \mathbb{R}^{d}$ are learnable tensors that enable the model to selectively emphasize relevant interaction patterns while suppressing noise. The resulting $\mathbf{K}_{z} \in \mathbb{R}^{d}$ serves as a compact yet comprehensive representation that effectively combines historical learning behaviors with future performance trends for the $z^{\textit{th}}$ attempt.

\subsubsection{Temporal Modeling and Performance Estimation} 
To effectively model the temporal dynamics inherent in the learning process, we employ an LSTM network to capture the sequential dependencies within the fused embeddings that combine historical learning sequences and FPTs. These fused embeddings are represented as $\mathbf{K} \in \mathbb{R}^{\bar{z} \times d}$, where $\bar{z}$ denotes the maximum number of attempts considered and $d$ represents the embedding dimension.

The LSTM processes the sequence of fused embeddings to generate a comprehensive temporal representation that encapsulates the evolution of a student's learning trajectory. Subsequently, we apply an MLP for dimensionality reduction to distill this temporal representation into a compact student knowledge state vector $\mathbf{e}_k \in \mathbb{R}^{d'}$, where $d'$ is the reduced dimension that captures the essential characteristics of the student's current knowledge state.

For performance prediction, we leverage another MLP to estimate the likelihood of student $s$ correctly answering a target question $\hat{o}$. This prediction is based on the interaction between the student's recent knowledge state $\mathbf{e}_k$ and the question embedding $\mathbf{e}_{\hat{o}}$. The prediction process is formulated as:
\begin{equation}
    \alpha_{\hat{o}}^{s} = \texttt{sigmoid}(\texttt{MLP}(\mathbf{e}_{k} \oplus \mathbf{e}_{\hat{o}})), 
    \label{eq:alpha}
\end{equation}
where $\oplus$ denotes the concatenation operation that combines the student's knowledge state with the question representation, and $\texttt{sigmoid}(\cdot)$ is the sigmoid activation function that maps the output to a probability range $[0,1]$. The resulting value $\alpha_{\hat{o}}^{s}$ represents the predicted probability that student $s$ correctly answers question $\hat{o}$, providing a quantitative measure of the student's expected performance on the given question.

\subsection{Objective Function}
In order to effectively train \textsf{FINER}, we utilize a loss function based on cross-entropy, which measures the discrepancy between the predicted probability $\alpha_{\hat{o}}^{s}$ that student $s$ correctly answers the target question $\hat{o}$, and the actual outcome $r$ (where $r=1$ if the answer is correct and $r=0$ otherwise). The loss function is defined as follows:
\begin{equation}
    \mathcal{L}(\theta) = -\sum_{(s,\hat{o},r)} \left[ r \log \alpha_{\hat{o}}^{s} + (1-r)\log(1-\alpha_{\hat{o}}^{s}) \right] + \lambda_{\theta}\|\theta\|^2,
\end{equation}
where $\theta$ represents the set of all learnable parameters within \textsf{FINER}. The term $\lambda_{\theta}\|\theta\|^2$ is a regularization component that helps prevent overfitting by penalizing large weights, where $\lambda_{\theta}$ is a hyperparameter that controls the strength of this regularization. The optimization of this objective function is performed using the Adam optimizer~\cite{kingma2014adam}. Detailed parameter settings and the specific configuration of the optimizer will be elaborated in the experimental section.

\setlength\tabcolsep{5.22pt}
\begin{table*}[!ht]
\centering
\small
\begin{tabular}{ c  cc  cc  cc  cc  cc  cc }
\Xhline{1pt}
\multirow{2}{*}{Method}  &    \multicolumn{2}{c}{\textbf{Assist09}}  &  \multicolumn{2}{c}{\textbf{Assist12}} &  \multicolumn{2}{c}{\textbf{Assist15}}  &  \multicolumn{2}{c}{\textbf{Algebra08}} & \multicolumn{2}{c}{\textbf{HDUOJ}}  & \multicolumn{2}{c}{\textbf{Junyi}} \\
\cline{2-13} 
          &  AUC    &  ACC   &  AUC  &  ACC  &  AUC   &  ACC   &   AUC   & ACC     &   AUC  & ACC     &  AUC  & ACC\\
\hline
DKT       &  75.49  &  72.42 & 70.72 & 72.90 &  72.66 &  74.93 &  67.92  &  84.36  &  91.64 &  85.85  & 84.83 & 79.19 \\
LSTMA     &  75.52  &  72.51 & 71.02 & 72.96 &  72.69 &  75.01 &  68.02  &  84.42  &  91.71 &  85.89  & 84.92 & 79.22 \\
DKVMN     &  74.75  &  71.92 & 69.40 & 72.25 &  72.23 &  74.96 &  67.20  &  84.43  &  90.59 &  85.01  & 82.41 & 76.92 \\
RKT       &  73.64  &  69.32 & 68.10 & 70.44 &  69.54 &  72.46 &  66.65  &  83.71  &  73.17 &  67.34  & 76.51 & 70.07 \\
SAKT      &  72.45  &  70.72 & 67.03 & 71.56 &  67.70 &  73.76 &  64.83  &  84.38  &  72.06 &  69.04  & 81.64 & 76.17 \\
AKT       &  78.51  &  73.91 & 71.08 & 73.13 &  73.23 &  75.20 &  68.18  &  84.65  &  92.62 &  87.14  & 85.41 & 78.75 \\
CoKT      &  78.56  &  74.12 & \underline{72.02} & \underline{74.26} &  73.46 &  75.22 &  68.21  &  84.87  &  92.68 &  87.36  & \underline{86.62} & \underline{81.12} \\ 
QIKT      &  \underline{78.82}  &  \underline{74.94} & 71.06 & 73.24 &  \underline{73.99} &  \underline{75.94} &  \underline{68.32}  &  \underline{85.34}  &  \underline{92.88} &  \underline{87.99}  & 86.24 & 81.03 \\
SimpleKT  &  77.46  &  73.21 & 70.48 & 72.83 &  72.83 &  74.73 &  67.88  &  84.12  &  91.01 &  86.74  & 84.52 & 77.45 \\
SparseKT  &  77.41  &  72.84 & 69.46 & 71.86 &  71.66 &  74.08 &  67.04  &  83.96  &  90.33 &  86.23  & 83.93 & 76.99 \\
\hline
\textsf{FINER}  & \textbf{83.23} & \textbf{77.46} & \textbf{83.26} & \textbf{77.95} & \textbf{83.35} & \textbf{79.64} & \textbf{95.20} & \textbf{92.70} & \textbf{93.65} & \textbf{89.04} & \textbf{93.86} & \textbf{85.85}\\ 
\hline
{Gain}     &  {20.82\%}     & {13.51\%} & {40.17 \%} & {14.34 \%} &  {35.99\%} &  {15.38\%}  &  {84.85\%} &  {50.20\%} &  {10.81\%} &  {8.74\%}  & 54.11 \%  & 25.05 \% \\
\Xhline{1pt}
\end{tabular}
\vspace{+2mm}
\caption{AUC (\%) and ACC (\%) of all methods, where bold indicates the best performance, underlining indicates the second-best performance and ``Gain'' denotes the performance gain of \textsf{FINER} over the best baseline.} 
\label{tab:CompareWithBaseline-new}
\end{table*}
\setlength\tabcolsep{1.2pt}
\begin{table}[!t]
\small
\centering
\begin{tabular}{ c  c  c  c  c  c  c  c }
\Xhline{1pt}
 Method              & Assist09   & Assist12 & Assist15 &  Algebra08 & HDUOJ  & Junyi  \\
 \hline
 DKT                   & 227 & 1,887 &   1,812 &   699 &  3,984 & 1,813\\
 LSTMA                & 245 & 2,503 &   2,364 &   739 &  4,718 & 2,099\\
 DKVMN              & 1,301 & 2,518 &  2,224 &   1,043 & 6,682 & 2,833\\
 RKT                    & \underline{223} & 1,916 &   \underline{1,739} &   \underline{658} & \underline{3,920} & 1,828\\
 SAKT                  & 229  & 1,861 &   1,775 &   688 &  4,108 & \underline{1,796}\\
 AKT                    & 1,092 & 1,896 &   1,804 &  909 &  6,830  & 3,968\\
 CoKT                  & 319 & 2,536 &   2,481 &   972 &  7,016  & 4,419\\
 QIKT                   & 262 & 1,847 &   1,792 &   784 &  4,298    & 2,227\\
 SimpleKT           & 226 & \underline{1,803} &   1,785 &   705 &  4,348    & 1,905\\
 SparseKT           & 304 & 1,914 &   1,889 &   915 &  6,878  & 2,147\\
 \hline
LPTrie Build        & 36  &  296   &  130  &  50  &  703  & 225  \\
FPT Search          & 19  &  186   &  182  & 136  &  68   & 147  \\
DL Module           & 158 &  1,285 & 1,367 & 452  & 2,988 & 1,399\\
\textsf{FINER}      & \textbf{214} &  \textbf{1,767} & \textbf{1,679} & \textbf{638}  & \textbf{3,759} & \textbf{1,771} \\
 \hline 
     Gain           &  4.04 \% &  2.00 \% &  3.45 \%  &  3.04 \%  &  4.11 \% & 1.39 \% \\
\Xhline{1pt}
\end{tabular}
\vspace{+1mm}
\caption{The training time (second) for all methods, where bold indicates the best performance, underlining indicates the second-best performance and ``Gain'' denotes the performance gain of \textsf{FINER} over the best baseline.}
\label{tab:CompareWithBaseline-time-2}
\vspace{+2mm}
\end{table}
\section{Experiments} \label{sec: Experiments}
\subsection{Experimental Setup}\label{app:datasets}
\subsubsection{Datasets}
We utilize six real datasets. The ASSISTments2009, ASSISTments2012, and ASSISTments2015  datasets \cite{abdelrahman_knowledge_2019} contain 325,637, 2,530,080, and 683,801 learning cells, respectively. Each features distinct sets of mathematical question types: 110 for both ASSISTments2009 and ASSISTments2015; 198 for ASSISTments2012, serving 4,151, 28,834, and 19,840 students, respectively. 

The Algebra08 dataset \cite{DBLP:conf/www/0001L0H00023} includes 1,048,575 learning cells from 424 questions answered by 247 students between September 19, 2008, and March 12, 2009. The Junyi dataset is sourced from Junyi Academy, a prominent Chinese e-learning platform. Following existing research \cite{xu2023learning}, we selected 1,000 learners who have the highest number of question-answering records from the log data, finally, it comprises 5,436,816 learning cells across 715 distinct questions. 

We also use a dataset, named HDUOJ, which is collected from \url{http://acm.hdu.edu.cn}. It includes 15,087,568 learning cells related to 5,320 questions from 137,374 students. This dataset provides a unique environment for examining knowledge tracing methods as it involves students learning independently without teacher intervention.

\subsubsection{Baselines and Performance Metrics}
We compare $\textsf{FINER}$ with ten state-of-the-art deep learning KT methods: 
three RNN-based KT methods: DKT \cite{DBLP:conf/nips/PiechBHGSGS15}, LSTMA \cite{DBLP:conf/aaai/SuLLHYCDWH18}, and QIKT \cite{DBLP:conf/aaai/00060HL023};
one MANN-based KT method: DKVMN \cite{DBLP:conf/www/ZhangSKY17}; five attention-based KT methods: RKT \cite{DBLP:conf/cikm/PandeyS20}, SAKT \cite{DBLP:conf/edm/PandeyK19}, AKT \cite{DBLP:conf/kdd/0001HL20}, SimpleKT \cite{DBLP:conf/iclr/0001L0H023}, and SparseKT \cite{huang2023towards}~(see Section~\ref{sec: related work}); and CoKT \cite{long_improving_2022} which utilizes collaboration information to enhance knowledge tracing.
We use three metrics: the area under the curve (AUC) \cite{abdelrahman_knowledge_2019}, accuracy (ACC), and training time. Specifically, for \textsf{FINER}, we include the time spent on building the Learning Pattern Trie and retrieving FPTs from it in the training time. Moreover, we report the runtime of the two processes separately
 (see Table~\ref{tab:CompareWithBaseline-time-2}).

\subsubsection{Experimental Setting}
We conduct 5-fold cross-validation on all datasets for each method. In each fold, we allocate 20\% of the student learning sequences for testing, another 20\% for validation, and the remaining 60\% for training. To enhance computational efficiency, we use the validation set in each fold for early stopping and parameter tuning. For sequences exceeding 200 learning cells, we follow reference~\cite{DBLP:conf/nips/PiechBHGSGS15} to split them into shorter sequences. 

\subsubsection{Hyperparameter Settings}
We explore weight decays and learning rates in the range $\{5e-3, 1e-3, 5e-4, 1e-4, 5e-5, 1e-5\}$ for all methods. We tune hyperparameters $\bar{i}$ and $\bar{z}$ in the range $1$ to $5$ while keeping other hyperparameters fixed, and we finally set $\bar{i} = 2$ and $\bar{z} = 2$ as defaults.
All results are averaged over experiments conducted using five different random seeds. And we publish the hyperparameter of \textsf{FINER} in \url{https://github.com/hyLiu1994/FINER}. 

\vspace{-3mm}
\subsection{Comparison}


Tables~\ref{tab:CompareWithBaseline-new} and~\ref{tab:CompareWithBaseline-time-2} report the accuracy and efficiency results, respectively. The performance improvement of \textsf{FINER} over the best-performing baseline is evaluated by the relative reduction in error rate, computed as:
\begin{equation}
\begin{aligned}
Gain &= \frac{\textsf{error}_{\text{baseline}} - \textsf{error}_{\text{FINER}}}{\textsf{error}_{\text{baseline}}}\\
  & = \frac{(1 - \textsf{score}_{\text{FINER}}) - (1 - \textsf{score}_{\text{baseline}})}{1 - \textsf{score}_{\text{baseline}}}\\
  & = \frac{\textsf{score}_{\text{FINER}} - \textsf{score}_{\text{baseline}}}{1 - \textsf{score}_{\text{baseline}}},
\end{aligned}
\end{equation}
where $\textsf{score}_{\text{FINER}}$ and $\textsf{score}_{\text{baseline}}$ denote the AUC or ACC scores achieved by \textsf{FINER} and the best baseline method, respectively.

\subsubsection{Accuracy study}  
Analysis of Table~\ref{tab:CompareWithBaseline-new} reveals two key findings:, we observe that: (i) \textsf{FINER} consistently outperforms all baselines across all datasets, achieving significant performance gains ranging from 8.74\% to 84.85\%; (ii) The performance gains vary across datasets (HDUOJ $<$ Assist15 $<$ Assist09 $<$ Assist12 $<$ Junyi $<$ Algebra08), with \textsf{FINER} achieving the highest AUC (95.20\%) and ACC (92.70\%) on Algebra08, which has the most learning cells per student. This is because the more learning cells a student has, the more learning scenarios the student experiences, and the more probability the student has to encounter correlation conflicts.


\setlength\tabcolsep{5.5pt}
\begin{table*}[!ht]
\centering
\footnotesize
\small
\setlength{\tabcolsep}{3.5pt}
\begin{tabular}{ c  cc  cc  cc  cc  cc  cc }
\Xhline{1pt}
\multirow{2}{*}{Method}  &    \multicolumn{2}{c}{\textbf{Assist09}}  &  \multicolumn{2}{c}{\textbf{Assist12}} &  \multicolumn{2}{c}{\textbf{Assist15}}  &  \multicolumn{2}{c}{\textbf{Algebra08}} & \multicolumn{2}{c}{\textbf{HDUOJ}}  & \multicolumn{2}{c}{\textbf{Junyi}} \\
\cline{2-13} 
         &  \makecell{11 $\rightarrow$ 0\\(7.05\%)}  &   \makecell{11 $\rightarrow$ 1 \\ (17.04\%)}  &   \makecell{11 $\rightarrow$ 0\\(8.44\%)}  &   \makecell{11 $\rightarrow$ 1\\(23.35\%)}  &   \makecell{11 $\rightarrow$ 0 \\ (10.15\%)}  &   \makecell{11 $\rightarrow$ 1 \\ (26.22\%)}  &    \makecell{11 $\rightarrow$ 0 \\ (8.16\%)}  &  \makecell{11 $\rightarrow$ 1 \\ (63.16\%)} &    \makecell{11 $\rightarrow$ 0 \\ (5.20\%)}  &  \makecell{11 $\rightarrow$ 1\\(5.87\%)} &    \makecell{11 $\rightarrow$ 0\\(6.13\%)}  &  \makecell{11 $\rightarrow$ 1\\(36.96\%)}\\
\hline
DKT       &  7.08  &  98.42 & 0.96 & \textbf{99.77}  &  1.63 &  98.83 &  0.01  &  99.99  &  86.99 &  \textbf{34.82}  & 0.47 & \textbf{99.94} \\
LSTMA     &  0.10  &  \textbf{99.98} & 0.50 & 99.75  &  0.34 &  \textbf{99.88} &  0.01  &  99.99  &  85.45 &  33.96  & 1.09 & 99.69 \\
DKVMN     &  6.65  &  97.49 & 4.84 & 98.36  &  4.25 &  98.95 &  0.10  &  99.95  &  86.24 &  34.25  & 15.16 & 93.90 \\
RKT       &  14.27  &  94.80 & 6.55 & 97.58 &  5.80 &  99.20 &  0.00  &  \textbf{100.00}  &  85.85 &  34.15  & 22.72 & 94.02 \\
SAKT      &  2.92  &  98.98 & 3.91 & 99.48 &  1.39 &  98.86 &  0.02  &  99.97  &  85.92 &  34.22  & 0.37 & 99.93 \\
AKT       &  2.50  &  99.35 & 2.31 & 98.41 &  1.38 &  99.84 &  0.00  &  \textbf{100.00} & 86.54 &  34.45 & 1.25 & 99.81 \\
CoKT      &  13.96  &  96.82 & 6.53 & 98.60 &  4.78 &  99.22 &  0.00  &  \textbf{100.00}  &  86.82 &  34.56  & 18.64 & 96.56 \\ 
QIKT      &  3.22  &  99.18 & 3.91  &  99.58 &  1.43 &  98.94 &  0.01  &  99.98  &  86.75 &  34.52  & 0.89 & 99.84 \\
SimpleKT  &  3.16  &  99.14 & 3.88  &  99.55 &  1.40 &  98.90 & 0.01 &  99.97 &  86.48 &  34.38  & 0.86 & 99.83 \\
SparseKT  & 3.20 & 99.16 &  3.90  &  99.56 & 1.42 &  98.91 &  0.01 & 99.98 &  86.65 & 34.48  & 0.92 & 99.79 \\
\hline
\textsf{FINER}  & \textbf{98.79} & 98.37 & \textbf{99.97} & 99.51 & \textbf{99.73} & 98.07 & \textbf{99.98} & 99.15 & \textbf{99.93} & 27.62 & \textbf{99.08} & 99.06\\ 
\Xhline{1pt}
\end{tabular}
\caption{ACC (\%) of all methods for the third attempt, where ``11 $\rightarrow$ 0'' and ``11 $\rightarrow$ 1'' represent sequences where students answered the same question correctly twice (11) followed by either an incorrect (0) or correct (1) attempt on that same question. The percentages in parentheses show the proportion of these patterns in each dataset.}
\label{tab:correlation-conflict-new}
\end{table*}
\setlength\tabcolsep{8pt}
\begin{table*}[!ht]
\centering
\small
\begin{tabular}{ c  cc  cc  cc  cc  cc  cc }
\Xhline{1pt}
\multirow{2}{*}{Scalability}  &    \multicolumn{2}{c}{\textbf{Assist09}}  &  \multicolumn{2}{c}{\textbf{Assist12}}  &  \multicolumn{2}{c}{\textbf{Assist15}}  &  \multicolumn{2}{c}{\textbf{Algebra08}} & \multicolumn{2}{c}{\textbf{HDUOJ}} & \multicolumn{2}{c}{\textbf{Junyi}} \\
\cline{2-13} 
         &  AUC  &  ACC  &  AUC  &  ACC  &  AUC  &  ACC  &   AUC  & ACC &   AUC  & ACC &   AUC  & ACC\\
\hline
20 \%  & 80.59 & 73.67 & 80.35 & 75.05 & 75.63 & 73.38 & 93.65 & 90.97 & 76.86 & 72.76 & 87.95 & 77.88 \\
40 \%  & 82.02 & 75.52 & 82.41 & 76.74 & 79.26 & 75.94 & 94.22 & 91.54 & 84.88 & 80.90 & 93.08 & 84.85 \\ 
60 \%  & 82.67 & 76.32 & 82.80 & 77.18 & 81.47 & 77.46 & 94.86 & 92.16 & 89.45 & 84.32 & 93.38 & 84.76 \\ 
80 \%  & 83.06 & 76.94 & 82.81 & 77.72 & 82.86 & 79.34 & 95.12 & 92.29 & 91.06 & 85.86 & 93.73 & 85.77 \\ 
100 \% & \textbf{83.23} & \textbf{77.46} & \textbf{83.26} & \textbf{77.95} & \textbf{83.35} & \textbf{79.64} & \textbf{95.20} & \textbf{92.70} & \textbf{93.65} & \textbf{89.04} & \textbf{93.86} & \textbf{85.85} \\ 
\Xhline{1pt}
\end{tabular}
\caption{The AUC (\%) and ACC (\%) of FINER across various data sizes on six datasets.}
\label{tab:Scalability-FINER}
\end{table*}

\setlength\tabcolsep{1.2pt}
\begin{table}[!t]
\small
\centering
\begin{tabular}{ c  c  c  c  c  c  c  c }
\Xhline{1pt}
 Scalability & Assist09  & Assist12  & Assist15 &  Algebra08 & HDUOJ  & Junyi  \\
 \hline
 20 \%       &  43   &  435  & 340   & 134   & 796    & 243\\
 40 \%       &  83   &  860  & 681   & 262   & 1,597  & 529\\
 60 \%       &  125  & 1,192 & 1,012 & 388   & 2,308  & 803\\
 80 \%       &  167  & 1,530 & 1,359 & 522   & 3,065  & 1,105\\
 100 \%      &  214  & 1,767 & 1,679 & 638   & 3,759  & 1,771 \\
\Xhline{1pt}
\end{tabular}
\caption{The training time (second) of FINER across various data sizes on six datasets.}
\label{tab:Scalability-Finer-time}
\end{table}

\setlength\tabcolsep{9pt}
\begin{table*}[!t]
\small
\centering{
\begin{tabular}{ c  cc  cc  cc  cc cc cc}
\Xhline{1pt}
\multirow{2}{*}{$\bar{i}$}  &    \multicolumn{2}{c}{\textbf{Assist09}}  &  \multicolumn{2}{c}{\textbf{Assist12}} &  \multicolumn{2}{c}{\textbf{Assist15}}  &  \multicolumn{2}{c}{\textbf{Algebra08}} & \multicolumn{2}{c}{\textbf{HDUOJ}}  & \multicolumn{2}{c}{\textbf{Junyi}} \\
\cline{2-13} 
         &  AUC  &  ACC  &  AUC  &  ACC  &  AUC  &  ACC  &   AUC  & ACC &   AUC  & ACC  &   AUC  & ACC  \\
\hline
$1$        &  78.32 &  74.26 & 81.29 & 76.66 &  71.78 &  67.54 &  68.24 &  64.01 &  72.98 &  68.84  & 85.09 & 79.21\\
$2$        & \textbf{83.23} & \textbf{77.46} & \textbf{83.26} & \textbf{77.95} & \textbf{83.35} & \textbf{79.64} & {95.20} & {92.70} & \textbf{93.65} & \textbf{89.04}  & \textbf{93.86} & \textbf{85.85} \\
$3$        &  82.95 &  76.92 & 82.03 & 76.51 &  80.45 &  76.57 &  95.32 &  92.81 &  85.45 &  80.75 & 92.77 & 82.93\\
$4$        &  82.10 &  75.28 & 81.89 & 76.36 &  74.05 &  70.10 &  \textbf{95.45} &  \textbf{92.96} &  79.24 &  74.43 & 89.80 & 78.80\\ 
$5$        &  81.27 &  74.96 & 82.53 & 77.27 &  71.24 &  65.97 &  95.41 &  92.79 &  73.57 &  69.72  & 87.87 & 75.57\\
\Xhline{1 pt}
\end{tabular}
\caption{AUC (\%) and ACC (\%) of \textsf{FINER} when varying $\bar{i}$, where bold indicates the best performance.}
\label{table:Np}
\vspace{+4mm}

\begin{tabular}{ c  cc  cc  cc  cc  cc  cc}
\Xhline{1pt}
\multirow{2}{*}{$\bar{z}$}  &    \multicolumn{2}{c}{\textbf{Assist09}}  &  \multicolumn{2}{c}{\textbf{Assist12}} &  \multicolumn{2}{c}{\textbf{Assist15}}  &  \multicolumn{2}{c}{\textbf{Algebra08}} & \multicolumn{2}{c}{\textbf{HDUOJ}}  & \multicolumn{2}{c}{\textbf{Junyi}} \\
\cline{2-13} 
&  AUC  &  ACC  &  AUC  &  ACC  &  AUC  &  ACC  &   AUC  & ACC &   AUC &  ACC  &  AUC  &  ACC  \\
\hline
$1$            &  81.18 &  76.12 & 71.34 & 72.89 &  81.11 &  77.84 &  92.99 &  89.79 &  85.93 &  81.88  & 90.71  &  82.74 \\
$2$            & \textbf{83.23} & \textbf{77.46} & \textbf{83.26} & \textbf{77.95}& \textbf{83.35} & \textbf{79.64} & \textbf{95.20} & \textbf{92.70} & \textbf{93.65} & \textbf{89.04}  & \textbf{95.06} & \textbf{87.13} \\
$3$            &  82.36 &  77.26 & 77.72 & 71.82 &  83.27 &  79.53 &  94.91 &  92.08 &  93.55 &  87.95 & 94.77 & 87.09 \\
$4$            &  82.22 &  77.08 & 73.60 & 66.62 &  83.23 &  79.49 &  94.27 &  91.70 &  92.98 &  86.29 & 94.59 & 87.11 \\
$5$            &  82.04 &  76.96 & 71.34 & 61.63 &  83.22 &  79.43 &  93.34 &  88.67 &  93.02 &  86.68 & 93.86 & 85.85 \\
\Xhline{1 pt}
\end{tabular}
\caption{AUC (\%) and ACC (\%) of \textsf{FINER} when varying $\bar{z}$, where bold indicates the best performance.} 
\label{table:Nf}
}
\end{table*}

\setlength\tabcolsep{4pt}
\begin{table*}[!t]
\small
\centering{
\begin{tabular}{ c  cc cc  cc  cc  cc  cc}
\Xhline{1pt}
\multirow{3}{*}{Metric}  &    \multicolumn{2}{c}{\textbf{Assist09}}  &  \multicolumn{2}{c}{\textbf{Assist12}} &  \multicolumn{2}{c}{\textbf{Assist15}}  &  \multicolumn{2}{c}{\textbf{Algebra08}} & \multicolumn{2}{c}{\textbf{HDUOJ}} & \multicolumn{2}{c}{\textbf{Junyi}} \\
\cline{2-13} 
         &  \textsf{FINER}  &  \textsf{FINER-S}  &  \textsf{FINER}  &  \textsf{FINER-S}  &  \textsf{FINER}  &  \textsf{FINER-S}  &   \textsf{FINER}  & \textsf{FINER-S} &   \textsf{FINER}  & \textsf{FINER-S} &   \textsf{FINER}  & \textsf{FINER-S} \\
\hline
AUC     & \textbf{83.23} & 82.67 & \textbf{83.26} & 82.50 & \textbf{83.35} & 82.61 & \textbf{95.20} & 94.95 & \textbf{93.65} & 93.40 & \textbf{93.86} & 93.64\\
ACC     & \textbf{77.46} & 76.94 & \textbf{77.95} & 77.06 & \textbf{79.64} & 78.86 & \textbf{92.70} & 92.45 & \textbf{89.04} & 88.88 & \textbf{85.85} & 85.55\\
Time    &    215     &     \textbf{203}    & 1,767 & \textbf{1,738} &   1,679    &   \textbf{1,641}   &   638   &    \textbf{621}   &    3,759    &  \textbf{3,732}   & 1,771 & \textbf{1,743} \\
\Xhline{1 pt}
\end{tabular}
\caption{AUC (\%), ACC (\%) and training time (second) of \textsf{FINER} and \textsf{FINER-S}, where \textsf{FINER-S} represents \textsf{{FINER}} without the similarity-aware attention mechanism.}
\label{tab:AblationOfModuleII}}
\end{table*}

\begin{table*}[!t]
\small
\centering{
\begin{tabular}{ c  ccc  ccc  ccc  ccc }
\Xhline{1pt}
\multirow{2}{*}{Metric}  &    \multicolumn{3}{c}{\textbf{Assist09-60}}  &  \multicolumn{3}{c}{\textbf{Assist09-70}}  &  \multicolumn{3}{c}{\textbf{Assist09-80}} & \multicolumn{3}{c}{\textbf{Assist09-90}} \\
\cline{2-13} 
 &  \textsf{FINER}  &  \textsf{FINER-K} &  \textbf{Gain} &  \textsf{FINER}  &  \textsf{FINER-K} &  \textbf{Gain} &  \textsf{FINER}  &  \textsf{FINER-K} &  \textbf{Gain} &  \textsf{FINER}  & \textsf{FINER-K} &  \textbf{Gain} \\ 
\hline
AUC  &  71.34  &   71.34   &  0.00\%   &  77.14  &   77.14 &  0.00\%  &   77.52  &  77.52  & 0.00\% & 77.58  &   77.58  &  0.00\%   \\ 
ACC  &  70.02  &   70.02   &  0.00\%   &  72.43  &   72.43 &  0.00\%  &   70.97  &  70.97  & 0.00\% &  72.93  &   72.93  &  0.00\%   \\ 
Time &  8.42  &   1,445   &   99.42\%  &  14.55  &   2,138 &  99.32\% &   14.98  &  2,886  & 99.48\% &15.83  &   3,528  &   99.55\%  \\ 
\Xhline{1 pt}
\end{tabular}
\caption{AUC (\%), ACC (\%) and training time (second) of \textsf{FINER} and \textsf{FINER-K}, where \textsf{FINER-K} represents \textsf{{FINER}} uses KMP algorithm instead of LPTrie, and ``Gain'' denotes the performance gain of \textsf{FINER} over \textsf{FINER-K}.
}
\label{tab:AblationOfModuleI}}
\end{table*}

\subsubsection{Efficiency study}
Table~\ref{tab:CompareWithBaseline-time-2} presents the training times of all methods, where LPTrie stands for \textbf{\underline{L}}earning \textbf{\underline{P}}attern \textbf{\underline{Trie}}. In contrast to the baselines, whose time costs are only for training the model, the time cost of \textsf{FINER} is divided into three parts: constructing the LPTrie (denoted as ``LPTrie Build''), retrieving FPTs from the LPTrie (denoted as ``FPT Search''), and training the DL Module (denoted as ``DL Module''), where DL Module represents the Aggregation and Fusion Modules. Despite these three parts, \textsf{FINER} demonstrates efficiency improvements of 1.39\% to 4.11\% across the six datasets. This is primarily attributed to two factors.

First, we provide efficient algorithms for LPTrie construction (cf. Algorithm~\ref{alg:BuildingLearningPatternExtractionAutomata}) and for retrieving FPTs from the LPTrie (Algorithm~\ref{alg:PotentialFuturePerformanceExtractor}). The time cost of building an LPTrie (from $36$s to $703$s) and the time cost of retrieving FPTs (from $19$s to $186$s) are considerably lower than the time cost (from $223$s to $3920$s) of training the baselines. 
Second, compared to the baselines, the DL Module, which incorporates FPTs, simplifies the modeling of student learning processes. This results in far fewer parameters needed --- approximately one-third of those required by the baselines --- without compromising prediction accuracy. Specifically, the training time of the DL Module (from 534s to 22,587s) is significantly less than those of the baselines (from 936s to 30,537s).

\subsection{Analysis of Correlation Conflict}
\label{sec:analysis-correlation-conflict}
We examine sequences of repeated attempts on the same question to validate the correlation conflict hypothesis. We focus on cases where students give two consecutive correct answers ("11"), followed by either a failure ("11 $\rightarrow$ 0") or success ("11 $\rightarrow$ 1"). By comparing how baselines and FINER handle these patterns, we evaluate their ability to address correlation conflicts.

Table~\ref{tab:correlation-conflict-new} presents the analysis across the six datasets, revealing several key findings:
The percentages in parentheses show that correlation conflicts (11 $\rightarrow$ 0) occur frequently, ranging from 5.20\% to 10.15\% of all cases. This indicates that even after two consecutive correct answers, students may fail on their next attempt. 
Next, the performance comparison shows that existing KT models struggle with these correlation conflicts. For sequences ending in failure (11 $\rightarrow$ 0), most baseline models achieve very low ACC scores (around 1\%--4\%) on datasets Assist09, Assist12, Assist15, Algebra08, and Junyi. This suggests they overwhelmingly predict success after seeing two correct answers, failing to identify scenarios where students might want to try different problem-solving strategies. 
As an exception, the baseline models show better, but still suboptimal, performance on the HDUOJ dataset, likely because it stems from an online programming platform where Scenario II (see Example~\ref{ex:motivation1}) is more common.
Finally, we see that \textsf{FINER} achieves consistently high ACC scores (98\%--99\%) for both "11 $\rightarrow$ 0" and "11 $\rightarrow$ 1" sequences across most of datasets. This is evidence of \textsf{FINER}'s ability to effectively differentiate between scenarios that appear similar based on recent performance but lead to different outcomes. The improvement is particularly dramatic for "11 $\rightarrow$ 0" sequences, where \textsf{FINER} outperforms baselines by large margins (e.g., by 14.27\% to 98.79\% on Assist09).

\subsection{Scalability} 
Tables~\ref{tab:Scalability-FINER} and \ref{tab:Scalability-Finer-time} present the AUC and ACC of the FINER model, as well as the corresponding training times across various data sizes. We can draw the following conclusions: 
(i) As the data scale increases from 20\% to 100\%, FINER's AUC and ACC improve significantly. For example, in Assist09 dataset, AUC increases from 84.59\% to 89.49\%, and ACC rises from 78.47\% to 82.17\%. Similarly, in HDUOJ dataset, AUC goes up from 76.86\% to 93.65\%, and ACC improves from 72.76\% to 89.04\%. This demonstrates that increasing the data scale significantly enhances model performance, and the smoothness of this improvement indicates the effectiveness of incorporating FPTs across different dataset sizes. 
(ii) The training time for FINER increases linearly with the data scale. For instance, in Assist09 dataset, the training time rises from 43 seconds to 214 seconds, and in HDUOJ dataset, it increases from 796 seconds to 3,759 seconds. This suggests that FINER's FPTs Search Module is highly efficient when scaling data, keeping the training time within an acceptable range, thereby proving its effectiveness.

\subsection{Parameter Study}

The two main parameters of \textsf{FINER} are $\bar{i}$,  the maximum length of learning patterns stored in and retrieved from the LPTrie, and $\bar{z}$, the maximum number of attempts considered after a pattern.

\noindent\textbf{{Impact of $\bar{i}$.}} Table~\ref{table:Nf} presents AUC and ACC results for \textsf{FINER} across the six datasets with $\bar{i}$ ranging from 1 to 5. Both AUC and ACC first grow and then drop as $\bar{i}$ increases. On the one hand, the narrow range of learning pattern lengths restricts \textsf{FINER} from extracting adequate FTPs essential for a precise understanding of student learning evolution. On the other hand, when \(\bar{i}\) increases, an overabundance of learning patterns in \textsf{FINER} may lead to overfitting of the DL Module. 

\noindent\textbf{{Impact of $\bar{z}$.}}
Table~\ref{table:Np} shows the AUC and ACC of \textsf{FINER} across the six datasets with $\bar{z}$ varying from $1$ to $5$.  The performance is modest at $\bar{z} = 1$, but improves with $\bar{z}>1$.  This is attributed to the inclusion of more subsequent attempts, enhancing the capture of FPTs and the understanding of learning evolution over time. The peak performance is usually observed at $\bar{z} = 2$, with a slight decrease when $\bar{z} > 2$. In latter cases, despite \textsf{FINER} acquiring more FPTs, their confidence reduces as these attempts are made further from the current moment. 

\subsection{Ablation Study}
Tables~\ref{tab:AblationOfModuleII} and~\ref{tab:AblationOfModuleI} present the results of an ablation study. \textsf{FINER-S} represents \textsf{{FINER}} without the similarity-aware attention mechanism, relying solely on FTP frequency for assigning attention weights.  Another variant, \textsf{FINER-K}, does not employ a LPTrie for storing and retrieving FPTs. Instead, it uses the KMP algorithm~\cite{anggreani2020knuth,sha2022chaotic} to match and extract FPTs directly from the historical learning sequences $\mathcal{X}$.

Table~\ref{tab:AblationOfModuleII} reveals that \textsf{FINER-S} has a slightly shorter training time than \textsf{FINER} across all datasets. However, \textsf{FINER-S} also has a lower accuracy than \textsf{FINER}. Specifically, \textsf{FINER} outperforms in both AUC, ranging from 3.78\% to 5.40\%, and ACC, from 1.44\% to 4.70\%, across five datasets, at the cost of an increase in training time from 0.72\% to 3.38\%. This  suggests that the similarity-aware strategy, despite adding some time overhead, effectively improves accuracy by capturing the interrelationship among learning patterns.

In Table~\ref{tab:AblationOfModuleI}, six smaller datasets are presented: Assist09-60, Assist09-70, Assist09-80, and Assist09-90. They are derived by sampling 60, 70, 80, and 90 learning sequences from Assist09. As the dataset size increases, the accuracy of \textsf{FINER-K} remains consistent with \textsf{FINER}. Notably, the efficiency of \textsf{FINER} improves significantly, by two orders of magnitude. 
This is attributed to the LPTrie, which effectively stores historical information and facilitates faster retrieval of FPTs without altering the data. 
The LPTrie assists \textsf{FINER} in significantly reducing the time complexity from \(O (\sum_{s\in S}(|X^s| + \bar{i} +\bar{z})) \) to $O(\bar{i})$ without impacting performance. 

\begin{figure}[!t]
  \centering
  \includegraphics[width=0.48\textwidth]{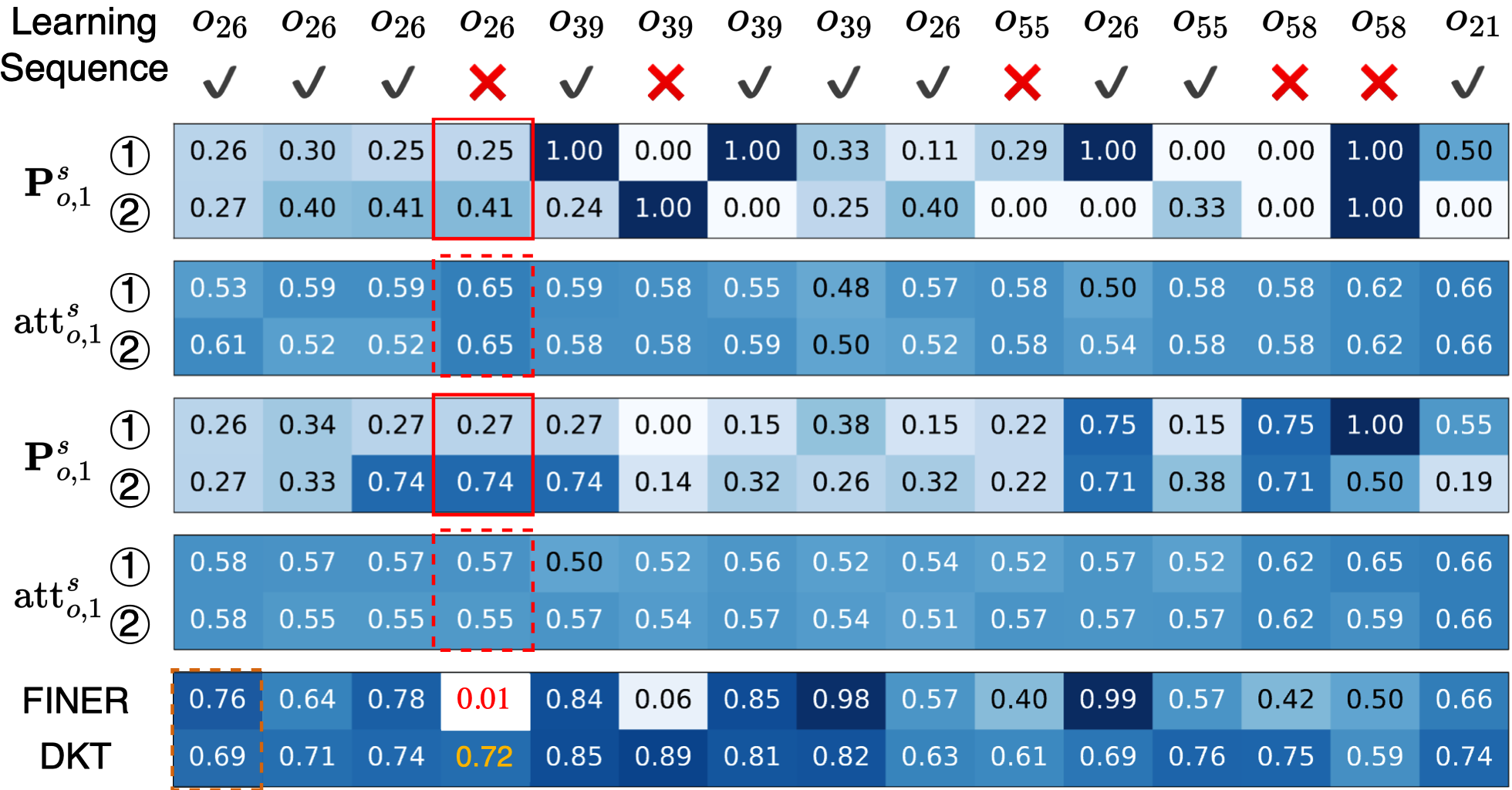}
  \caption{Visualization of KT, where (i) the top section shows a real learning sequence from Assis09; (ii) the middle section displays the ratio matrix of FPTs along with their attention weights; and (iii) the bottom section compares the predictions made by \textsf{FINER} and DKT.}
  \label{fig:VisualizingOfFPTs}
\end{figure}

\subsection{\textsf{FINER} Visualization}
In the top part of Fig.~\ref{fig:VisualizingOfFPTs}, we present some learning sequences $\langle (o_{26}, 1), (o_{26}, 1), ..., (o_{58}, 0), (o_{58}, 0)\rangle$ of a student $s$, extracted from Assist05. The target question is $o_{26}$ and both $\bar{i}$ and $\bar{z}$ are set to 2.

\noindent\textbf{{Effectiveness of exploring FPTs.}}  
Assuming that currently, $s$ has completed the learning sequence $X^s=\langle (o_{26}, 1), (o_{26}, 1), (o_{26}, 1) \rangle$ and \textsf{FINER} aims to predict the probability of $s$ correctly answering $o_{26}$ in the next attempt of $s$. We first extract two learning patterns from $X^s$: $v^s_1 = \langle (o_{26}, 1) \rangle$ and $v^s_2 = \langle (o_{26}, 1), (o_{26}, 1) \rangle$. Correspondingly, we identify two FPTs from the LPTrie: $(1, [0.25, 0.41],  [984, 984])$ and $(2, [0.27, 0.74],  [876, 798])$. As a result, $\mathbf{P}^s_{o_{26},1} = [0.25, 0.41]$ and $\mathbf{P}^s_{o_{26},2} = [0.27, 0.74]$ (highlighted in the red solid box), indicating that, based on ITS historical data, the frequencies of students correctly answering $o_{26}$ on the $1^{\textit{st}}$ and $2^{\textit{nd}}$ attempts of $s$ after $v^s_1$ are 25\% and 41\%, respectively. 
Next, we calculate the similarity matrix, which measures the similarity between two FPTs (cf. Section~\ref{sec:Similarity-aware_Attention_Mechanism}) and yields their corresponding attention weights $\textit{att}^s_{o_{26},1} = [0.65, 0.65]$ and $\textit{att}^s_{o_{26},2} = [0.57, 0.55]$ (highlighted in the red dashed box). The DL Module uses the weights to prioritize $\mathbf{P}^s_{o_{26},i}[z]$ in predicting student performance--higher values of $\textit{att}^s_{o_{26},i}[z]$ indicates more importance of $\mathbf{P}^s_{o_{26},i}[z]\,(z=\{0,1\})$.
Finally, we fuse the aggregated FPT embeddings $\mathbf{T}^s_{\hat{o}}$ with the embedding of $X^s$. This produces a predicted probability of 0.01 (indicated by the red digit).

The actual learning history of student $s$ shows that he answered $o_{26}$ incorrectly after completing $X^s$, aligning with \textsf{FINER}'s prediction. Despite the initial success of $s$ on $o_{26}$, there is a notable difference in the predicted probabilities: $\mathbf{P}^s_{o_{26},2}[1]=0.27$ versus $\mathbf{P}^s_{o_{26},2}[1]=0.74$. This gap suggests a significant change in performance on $o_{26}$ between the first and second attempts after experiencing $v_2$. This variation, potentially indicative of intensive or pre-exam practice (cf. Example~\ref{ex:motivation1}), is captured by $\textsf{FINER}$. In contrast, the DKT method, indicating a high probability (0.72, shown as an orange digit) of a correct answer post $X^s$, overlooks the learning context present in historical data and focuses solely on the historical learning sequence $X^s$ of $s$.

\noindent\textbf{Effectiveness of similarity-aware attention mechanism.}
The first column of Fig.~\ref{fig:VisualizingOfFPTs} shows that $\mathbf{P}^s_{o_{26},1}$ and $\mathbf{P}^s_{o_{26},2}$ are both [0.26, 0.27]. Consequently, we obtain attention weights $\textit{att}^s_{o_{26},1} = [0.53, 0.61]$ and $\textit{att}^s_{o_{26},2} = [0.58, 0.58]$, which are quite similar. This occurs despite $\omega^s_{o_{26},2} = [13509, 13333]$ being significantly higher than $\omega^s_{o_{26},1} = [4071, 3882]$.   This is because the proposed mechanism prioritizes the 100\% similarity between $\mathbf{P}^s_{o_{26},1}$ and $\mathbf{P}^s_{o_{26},2}$, which makes their confidence levels similar and diminishes the influence of frequency on attention weights. 
In contrast, in the fourth column in Fig.~\ref{fig:VisualizingOfFPTs}, 
$\mathbf{P}^s_{o_{26},1}$ and $\mathbf{P}^s_{o_{26},2}$ show different values: [0.25, 0.41] and [0.27, 0.74], respectively, a situation also seen in the third column of the figure. In these cases, the proposed mechanism does not affect the weights $\textit{att}^s_{o_{26},1}$ and $\textit{att}^s_{o_{26},2}$ significantly, and they are mainly influenced by the frequencies of FPTs. As a result, the attention weights in the third column are similar to those in the fourth column. The prediction results, highlighted in orange solid boxes, demonstrate the effectiveness of the proposed mechanism.

\section{Conclusions and Future Work}
\label{sec: Conclusion}
This paper proposes \textsf{FINER}, a novel knowledge tracing method that resolves correlation conflicts by integrating Follow-up Performance Trends (FPTs) with historical learning sequences.
\textsf{FINER} contains three key modules. The FPT Fetching Module's innovative use of a learning pattern trie significantly streamlines FPT retrieval. The Multiple-FPT Aggregation Module effectively identifies the confidence of varying lengths of learning patterns for enhanced aggregation. The Recent History Fusion Module offers a more comprehensive view of students' behaviors for improving prediction accuracy. Experiments on six real-world datasets show that \textsf{FINER} outperforms SOTA methods, improving prediction accuracy by 8.74\% to 84.85\%. 
In future research, it is of interest to explore how to achieve better performance in complex learning situations, such as those covered by the HDU dataset.

\section{Acknowledgment}
The research leading to the results presented in this paper has received funding from the European Union’s funded Project MobiSpaces under grant agreement no 101070279, and partial support from the National Natural Science Foundation of China (No. 62272093).

\bibliographystyle{IEEEtran}
\bibliography{FINER}
\vspace{-10mm}
\begin{IEEEbiography}[{\includegraphics[width=1in,height=1.25in,clip,keepaspectratio]{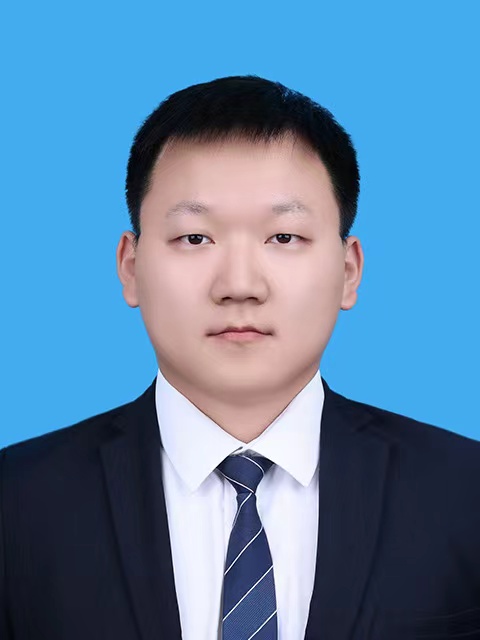}}]{Hengyu Liu} (Member, IEEE)
    received the Ph.D. degree in Computer Application Technology from the Northeastern University, Shenyang, China, in 2023. He is currently a Postdoc Researcher with the Department of Computer Science, Aalborg University, Aalborg, Denmark. His main research interests include machine learning, intelligent education, and spatial-temporal data management.
    \end{IEEEbiography}
    \vskip -1\baselineskip plus -1fil
    
    \begin{IEEEbiography}
    [{\includegraphics[width=1in,height=1.25in,clip,keepaspectratio]{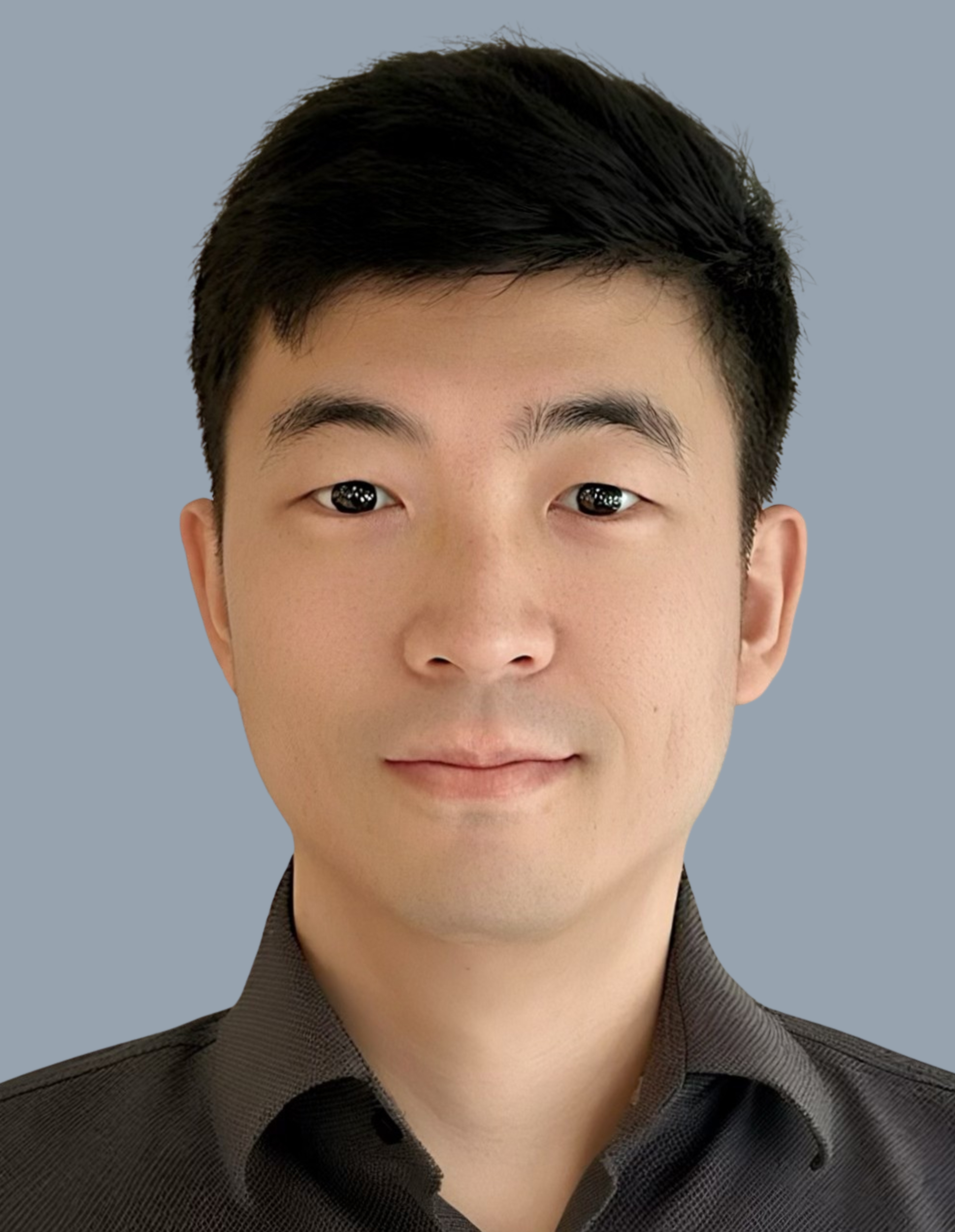}}]
    {Yushuai Li} (Senior Member, IEEE) received the Ph.D. degree in control theory and  control engineering from the Northeastern University, Shenyang, China, in 2019.  He is currently an Assistant Professor with the Department of Computer Science, Aalborg University, Aalborg, Denmark.  He received the Best Paper Awards from Journal of Modern Power Systems and Clean Energy (MPCE) in 2021, and ICCSIE in 2023, IEEE EI2 in 2024. He serves as Associate Editors for IEEE Transactions on Industrial Informatics, IEEE Transactions on Automation Science and Engineering, MPCE, and IEEE Systems, Man, and Cybernetics Magazine. His main research interests include machine learning, digital twin, and integrated energy and transportation systems.
    \end{IEEEbiography}
    
    \vskip -1\baselineskip plus -1fil
    \begin{IEEEbiography}[{\includegraphics[width=1in,height=1.25in,clip,keepaspectratio]{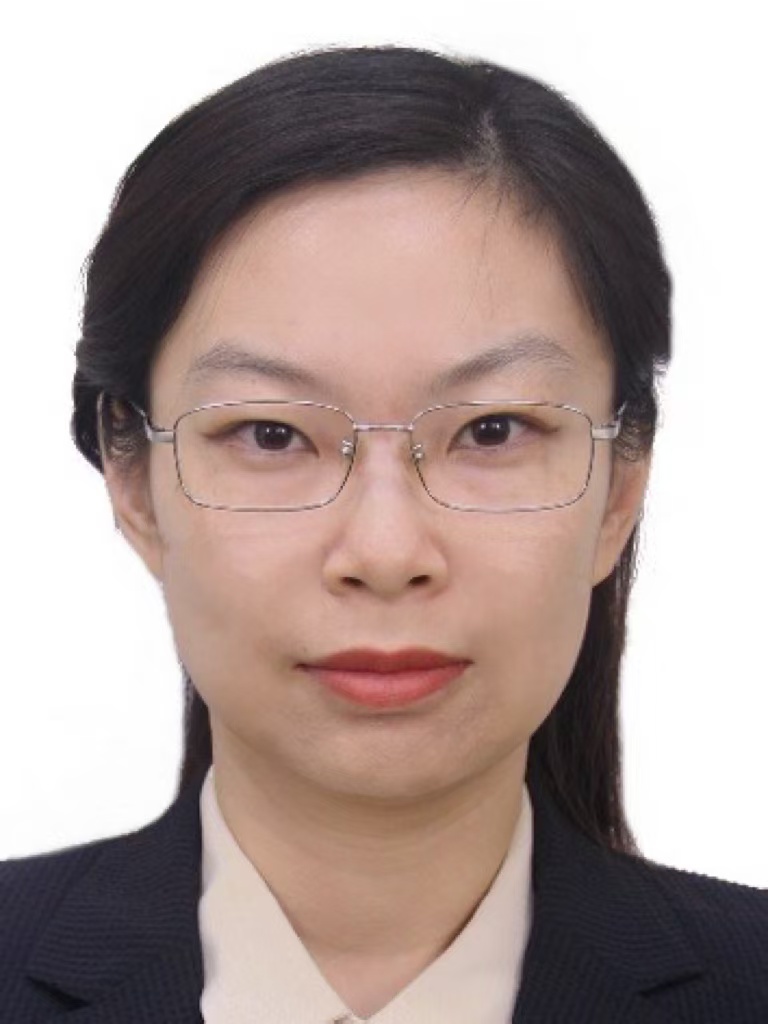}}]{Minghe Yu} received the BS degree in computer science and technology from Northeastern University, Shenyang, China, in 2012, and the PhD degree in computer science and technology from Tsinghua University, Beijing, China, in 2018. She is currently an associate professor with Software College, Northeastern University, Shenyang, China. Her research interests include database, information retrieval and intelligent education..
    \end{IEEEbiography}
    
    \vskip -1\baselineskip plus -1fil
    \begin{IEEEbiography}[{\includegraphics[width=1in,height=1.25in,clip,keepaspectratio]{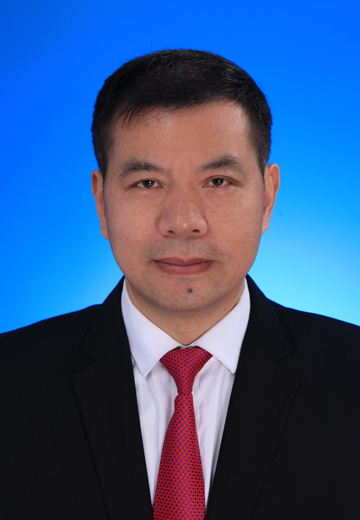}}]{Tiancheng Zhang} received a Ph.D degree in computer software and theory from Northeastern University (NEU) of China. He is currently an associate professor in the School of Computer Science and Engineering at NEU. His research interests include big data analysis, spatiotemporal data management, and deep learning.
    \end{IEEEbiography}
    
    \vskip -1\baselineskip plus -1fil
    \begin{IEEEbiography}[{\includegraphics[width=1in,height=1.25in,clip,keepaspectratio]{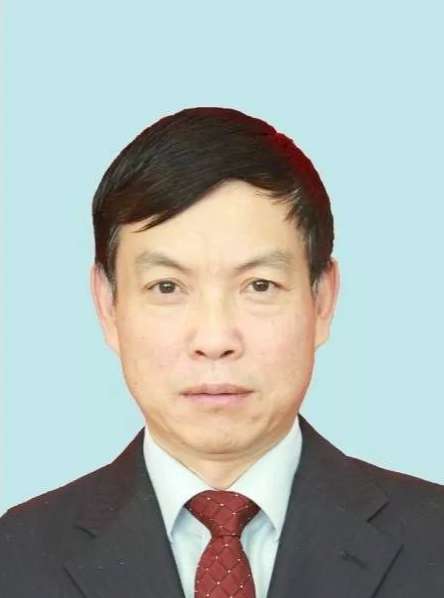}}]{GE YU} (Senior Member, IEEE) received  Ph.D. degree in computer science from Kyushu University, Japan, in 1996. He is currently a professor and a Ph.D. Supervisor at Northeastern University, China. His research interests include distributed and parallel databases, OLAP and data warehousing, data integration, and graph data management. He is a member of ACM and a Fellow of the China Computer Federation (CCF).
    \end{IEEEbiography}
    
    \vskip -1\baselineskip plus -1fil
    \begin{IEEEbiography}
    [{\includegraphics[width=1in,height=1.25in,clip,keepaspectratio]{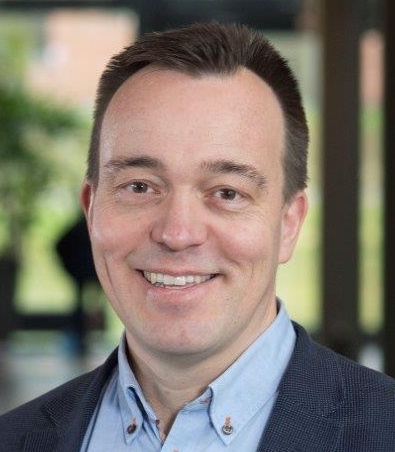}}]
    {Torben Bach Pedersen}(Senior Member, IEEE) is a professor with the Center for Data-Intensive Systems (Daisy), Aalborg University, Denmark. His research concerns Predictive, Prescriptive, and Extreme-Scale Data Analytics with Digital Energy as the main application area. He is an ACM Distinguished Scientist, an IEEE Computer Society Distinguished Contributor, a member of the Danish Academy of Technical Sciences, and holds an honorary doctorate from TU Dresden.
    \end{IEEEbiography}

    \vskip -1\baselineskip plus -1fil
    \begin{IEEEbiography}[{\includegraphics[width=1in,height=1.25in,clip,keepaspectratio]{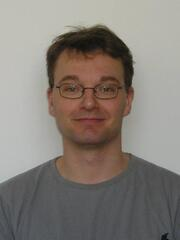}}]{Kristian Torp} is a Professor in the Department of Computer Science at Aalborg University, where he works within the Center for Data-Intensive Systems (Daisy). His research focuses on spatio-temporal data management, trajectory analytics, and intelligent transportation systems. He has published extensively in top-tier venues including ICDE, VLDB, SIGSPATIAL and other leading databases and data management conferences. He has been involved in numerous research projects addressing scalable data processing and urban mobility challenges, and collaborates closely with both industry and public-sector partners.
    \end{IEEEbiography}

    \vskip -1\baselineskip plus -1fil
    \begin{IEEEbiography}[{\includegraphics[width=1in,height=1.25in,clip,keepaspectratio]{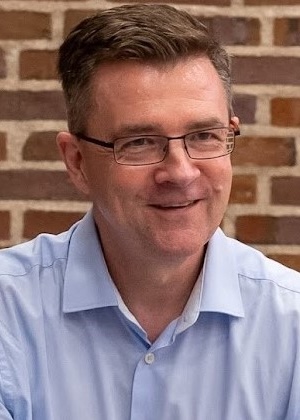}}] {Christian S. Jensen} (Fellow, IEEE) received the PhD degree from Aalborg University, in 1991, and the DrTechn degree from Aalborg University, in 2000. He is a professor with the Department of Computer Science, Aalborg University. His research concerns primarily temporal and spatiotemporal data management and analytics, including indexing and query processing, data mining, and machine learning.
    \end{IEEEbiography}

    \vskip -1\baselineskip plus -1fil
    \begin{IEEEbiography}[{\includegraphics[width=1in,height=1.25in,clip,keepaspectratio]{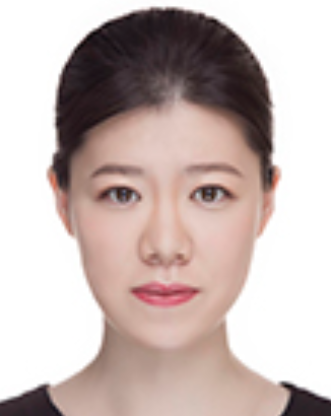}}] {Tianyi Li} (Member, IEEE) received the Ph.D. degree from Aalborg University, Denmark, in 2022. She is an Assistant Professor at the Department of Computer Science, Aalborg University. She received ICDE 2022 Best Paper Award.  She serves as an Associate Editor for IEEE Network and IEEE Transactions on Intelligent Vehicles. Her research concerns spatio-temporal data, intelligent transportation, machine learning, time series, and database technology.
    \end{IEEEbiography}
\end{document}